\documentclass[pra,twocolumn]{revtex4}

\usepackage{amsmath}
\usepackage{color}
\usepackage{dcolumn}
\usepackage[cp1251]{inputenc}
\usepackage[english,russian]{babel}
\usepackage{epsfig}
\usepackage{array}

\begin{document}

\title{Entanglement of qutrits and ququarts}

\author{M.V. Fedorov$^{1\,*}$, P.A. Volkov$^{1}$, Yu.M. Mikhailova$^{1,\,2}$,}
\affiliation{$^1$A.M.Prokhorov General Physics Institute, Russian Academy of Science, Moscow, Russia\\
$^2$Max-Planck-Institut $f\ddot{u}r\,$ Quantenoptik, Garching, Germany\\
$^*$e-mail: fedorov@gmail.com}

\author{S.S. Straupe, S.~P.~Kulik}
\affiliation{Faculty of Physics, M.V.Lomonosov Moscow State University, Moscow, Russia}

\date{\today}

\begin{abstract}

We investigate in a general form entanglement of biphoton qutrits and ququarts, i.e. states formed in the processes of collinear and, correspondingly, degenerate and non-degenerate Spontaneous Parametric Down-Conversion. Indistinguishability of photons and, for ququarts, joint presence of the frequency and and polarization entanglement are fully taken into account. In the case of qutrits the most general 3-parametric families of maximally entangled and non-entangled states are found, and anti-correlation of the degree of entanglement and polarization is shown to occur and to be characterized by a rather simple formula. Biphoton ququarts are shown to be two-qudits with the single-photon Hilbert space dimensionality $d=4$, which differs them significantly from the often used two-qubit model ($d=2$). New expressions for entanglement quantifiers of biphoton ququarts are derived and discussed. Rather simple procedures for a direct measurement of the degree of entanglement are described for both qutrits and ququarts.

\end{abstract}


\maketitle

\def\thesection{\arabic{section}}
\numberwithin{equation}{section}

\section{Introduction}

Optical qutrits and ququarts are promising objects of the modern quantum
information and quantum cryptography \cite{Peres-PRL,Gisin,Bogd-04,Bogd-04a,Bogd-06,Molotkov,Genovese}.
Formally, qutrits and ququarts are defined as superpositions of, correspondingly, three and four
basis states. In practice, most often, the basis states for qutrits and ququarts are formed by biphoton states arising in the processes of Spontaneous Parametric Down-Conversion (SPDC). For qutrits it is sufficient to use the   collinear degenerate SPDC processes, i.e., such processes in which wave vectors of two photons in a
SPDC pair are strictly parallel to each other and
frequencies are also given and equal to each other. For constructing ququarts, one has to use either the non-collinear frequency-degenerate or collinear but frequency-non-degenerate SPDC processes, i.e., processes in which either directions of wave vectors or frequencies of two photons in SPDC pairs differ from each other. Below we will refer both of these possibilities as ``non-degenerate$"$.
In theory, from the very beginning \cite{Klyshko-99}, biphoton qutrits were considered as arbitrary superpositions of
three Fock states, corresponding to three possibilities of distributing two indistinguishable photons of an SPDC pair in two polarization modes, horizontal ($H$) and vertical ($V$) ones. Each of these basis states is a direct product of two one-qubit single-photon states, and biphoton qutrits are two-qubit states. There were many works devoted to biphoton qutrits \cite{Bu-Che-Ka,Bu-Che-Kl,Bu-Che,Bu-Kri-Ku,Kri-Ku-Pe,Bogd-03,Klyachko}. Most of them study polarization properties of qutrits and much less entanglement. Giving a comprehensive picture of entanglement of qutrits is one of the goals of this paper. In particular, we will find and describe general families of maximally entangled and non-entangled qutrits, their entanglement quantifiers such as the Schmidt entanglement parameter, concurrence, and the subsystem entropy, relations between entanglement and polarization of qutrits, etc.

An important point to be discussed is the role of the symmetry of biphoton wave functions with respect to permutation of particles' variables. Existence of entanglement related to symmetry was realized by many authors rather long ago, \cite{Schliemann, Paskauskas, Peres, Zanardi, Lanyon} and so on. But a general attitude to the entanglement related to symmetry is rather sceptical, up to the opinion that this type of entanglement is unimportant and can be forgotten. In such approach all basis states of qutrits and ququarts would be non-entangled and the only reason for entanglement would be related to various choice of coefficients in superpositions of basis states. This is a configurational entanglement. But in reality, owing to symmetry, basis states of qutrits and ququarts are entangled (at least one basis states in the case of qutrits and all basis states of ququarts). This is a fundamental, unchangeable entanglement. In superpositions of basis states the symmetry-related and configuration entanglement exist together and each of them is so strongly built into a general picture of entanglement of qutrits and ququarts, and into their entanglement quantifiers, that it's impossible to separate these two types of entanglement and to take off the symmetry-related one without hurting significantly the general picture and results. Thus, all kinds of entanglement have to be taken into account together and none of them can be ``forgotten$"$.

In the case of  ququarts, the symmetry-related entanglement arises from both polarization and frequency (or angular) degrees of freedom. This makes the traditional two-qubit model of biphoton ququarts invalid. As we show, biphoton ququarts are two-qudit states with the dimensionality of the one-photon Hilbert space $d=4$ and dimensionality of the two-photon Hilbert space $D=d^2=16$. This makes ququarts significantly different from qutrits (where $d=2$ and $D=d^2=4$), and a series of new results on entanglement of ququarts is derived and discussed in Section 7. This new understanding of the physics of ququarts raises a question about changes  in applications of ququarts analyzed earlier in the frame of the two-qubit model when the latter is substituted by the qudit picture. We hope to return to such analysis elsewhere.

In addition to the above-mentioned entanglement quantifiers, we use widely the Schmidt decomposition of biphoton wave functions \cite{Grobe, Knight}. As known \cite{Rungta}, for pure biphoton states with the dimensionality of the one-photon Hilbert space $d$ the maximal amount of terms in the Schmidt decomposition equals $d$, and such states are unseparable. Only if all but one coefficients in the Schmidt decomposition are equal zero, and the remaining exceptional coefficient equals unit, the Schmidt decomposition is reduced to a single product of Schmidt modes, and such state is separable. We consider this criterion as the ultimate indication of whether states are separable or not.

\section{State vectors and wave functions of biphoton qutrits}

In the form of state vectors, purely polarization biphoton states (qutrits), are given by a superposition
\begin{equation}
 \label{state-vector}
 |\Psi\rangle=C_1|2_H\rangle+C_2|1_H,1_V\rangle+C_3|2_V\rangle,
\end{equation}
where the basis state vectors are given by
\begin{equation}
 \label{bas-stat-vect}
 |2_H\rangle=\frac{1}{\sqrt{2}}a_H^{\dag\,2}|0\rangle,\,|1_H,1_V\rangle=a_H^\dag
 a_V^\dag|0\rangle,\,|2_V\rangle=\frac{1}{\sqrt{2}}a_V^{\dag\,2}|0\rangle,
\end{equation}
$|0\rangle$ is the vacuum state vector, $a_H^\dag$ and
$a_V^\dag$ are the creation operators of photons in the modes with
horizontal and vertical polarizations (with given equal frequencies and given identical propagation directions).  $C_{1,2,3}$ are
arbitrary complex constants
$C_{1,2,3}=|C_{1,2,3}|e^{i\varphi_{_{1,2,3}}}$, obeying the
normalization condition
\begin{equation}
 \label{norm-qtr}
 |C_1|^2+|C_2|^2+|C_3|^2=1.
\end{equation}
Actually, as the total phase of the state vector
(\ref{state-vector}) or wave function (see below) does not affect
any measurable characteristics of qutrits, one of the phases
$\varphi_{_{1,2,3}}$, or a linear combination of phases, can be taken equal zero,
and, hence, the general form of the qutrit state
vector (\ref{state-vector}) is characterized by four independent
constants (e.g., $|C_1|$, $|C_3|$, $\varphi_1$, and
$\varphi_3$ with $\varphi_2=0$).

As qutrit (\ref{state-vector}) is a two-photon state, its polarization wave function depends on two variables.
A general rule of obtaining multipartite wave functions from state vectors is known pretty well in the quantum-field theory, and for bosons the corresponding formula has the form \cite{Shweber} (in slightly modified notations)
\begin{gather}
 \nonumber
 \Psi(x_1,\,x_2, ..., x_n)=\langle x_1,\,x_2, ..., x_n|n_1,\,n_2, ...,\,n_k\rangle \\
 \nonumber
 =\frac{1}{\sqrt{n_1!n_2!...n_k!n!}}\sum_PP\Big(g_{1}(x_1)g_1(x_2)...g_{1}(x_{n_1})\\
 \nonumber
 \times g_{2}(x_{n_1+1})g_{2}(x_{n_1+2})...g_{2}(x_{n_1+n_2})...\\
 \label{Schweber}
 \times g_{k}(x_{n_1+n_2+...+n_{k-1}+1})...g_{k}(x_n)\Big),
\end{gather}
where $x_1,\,x_2, ...,\,x_n$ are dynamical variables of identical boson particles, $g_j(x_i)$ are single-particle wave functions (of $j$-th modes and $i$-th variables), $P$ indicates all possible transpositions of variables $x_i$ in wave functions $g_j(x_i)$, $n_1,\,n_2, ...,\,n_k$ are numbers of particles in modes, $n$ is the total amount of particles in all modes, $k$ is the total amount of modes; for empty modes the corresponding single-particle wave functions have to be dropped.

In the case of qutrits we have two modes ($j=H\,{\rm or}\,V$) and two particles, $n=k=2$. The polarization variables of two photons can be denoted as $\sigma_1$ and $\sigma_2$. In terms of wave functions, the single-photon wave functions $g_j(x_i)$ are given by the  Kronecker symbols. Thus, the qutrit basis wave functions corresponding to the basis state vectors in Eq. (\ref{state-vector}) can be written as
\begin{gather}
 \label{HH}
 \Psi_{HH}(\sigma_1,\sigma_2)=\langle\sigma_1,\sigma_2|2_H\rangle=\delta_{\sigma_1,H}\delta_{\sigma_2,H},\\
 \nonumber
 \Psi_{HV}(\sigma_1,\sigma_2)=\langle\sigma_1,\sigma_2|1_H,1_V\rangle\\
 \label{HV}
 =\frac{1}{\sqrt{2}}\,\Big[\delta_{\sigma_1,H}\delta_{\sigma_2,V}+\delta_{\sigma_2,H}\delta_{\sigma_1,V}\Big],\\
 \label{VV}
 \Psi_{VV}(\sigma_1,\sigma_2)=\langle\sigma_1,\sigma_2|2_V\rangle=\delta_{\sigma_1,V}\delta_{\sigma_2,V}.
\end{gather}
The same basis wave functions can be written equivalently in the form of 2-row columns, which is more convenient for calculation of matrices
\begin{gather}
 \label{columns-HH}
 \Psi_{HH}=\left({1\atop 0}\right)_1\otimes\left({1\atop 0}\right)_2\equiv\left({1\atop 0}\atop{0\atop 0}\right),\\
 \label{columns-VV}
 \Psi_{VV}=\left({0\atop 1}\right)_1\otimes\left({0\atop 1}\right)_2\equiv\left({0\atop 0}\atop{0\atop 1}\right),
\end{gather}
\begin{gather}
 \nonumber
 \Psi_{HV}=\frac{1}{\sqrt{2}}\left[\left({1\atop 0}\right)_1\otimes\left({0\atop 1}\right)_2+\left({0\atop 1}\right)_1\otimes\left({1\atop 0}\right)_2\right]\\
 \label{columns-HV}
 \equiv\frac{1}{\sqrt{2}}\left[\left({0\atop 1}\atop{0\atop 0}\right)+\left({0\atop 0}\atop{1\atop 0}\right)\right]
 =\frac{1}{\sqrt{2}}\left({0\atop 1}\atop{1\atop 0}\right),
\end{gather}
where the upper and lower rows in two-row columns correspond to the horizontal and vertical polarizations and the indices 1 and 2 numerate indistinguishable photons.

In a general form the qutrit wave function corresponding to the state vector (\ref{state-vector}) is given by
\begin{equation}
 \label{qutrit-general}
 \Psi= C_1\Psi_{HH}+C_2\Psi_{HV}+C_3\Psi_{VV},
\end{equation}
where $\Psi_{HH}$, $\Psi_{HV}$, and $\Psi_{VV}$ can be taken either in the form (\ref{HH})-(\ref{VV}) or  (\ref{columns-HH})-(\ref{columns-HV}).

Alternatively, the same general qutrit wave function (\ref{qutrit-general}) can be presented in the form of an expansion in a series of Bell states
\begin{equation}
 \label{Bell-exp}
 \Psi= C_+\Phi^++C_2\Psi^++C_-\Phi^-,
\end{equation}
where
\begin{equation}
 \label{C-pm}
 C_\pm=\frac{C_1\pm C_3}{\sqrt{2}},
\end{equation}
$\Psi^+\equiv \Psi_2$ (\ref{columns-HV})
and
\begin{gather}
 \nonumber
 \Phi^\pm=\frac{1}{\sqrt{2}}\,\Big[\Psi_{HH}\pm\Psi_{VV}\Big]\\
 \nonumber
 = \frac{1}{\sqrt{2}}
 \left[\left({1\atop 0}\right)_1\otimes\left({1\atop 0}\right)_2
 \pm\left({0\atop 1}\right)_1\otimes\left({0\atop 1}\right)_2\right]\\
 \label{Phi+-}
 \equiv\frac{1}{\sqrt{2}}\left[\left({1\atop 0}\atop{0\atop 0}\right)\pm\left({0\atop 0}\atop{0\atop 1}\right)\right]
 =\frac{1}{\sqrt{2}}\left({1\atop 0}\atop{0\atop \pm 1}\right).
\end{gather}
$\Phi^\pm$ and $\Psi^+$  are the wave functions describing three Bell states. The fourth Bell state,
\begin{equation}
 \label{Bell-fourth}
 \Psi^-=\frac{1}{\sqrt{2}}\,
 \left[\left({1\atop 0}\right)_1\otimes\left({0\atop 1}\right)_2
 -\left({0\atop 1}\right)_1\otimes\left({1\atop 0}\right)_2\right],
\end{equation}
does not and cannot arise in the expansion (\ref{Bell-exp}) because $\Psi^-$ (\ref{Bell-fourth}) is antisymmetric with respect to the variable transposition $1\rightleftharpoons 2$, whereas all biphoton wave functions have to be symmetric. Nevertheless, in principle, the antisymmetric Bell state (\ref{Bell-fourth}) can be included into the expansion (\ref{Bell-exp}) but obligatory with the zero coefficient:
\begin{equation}
 \label{Bell-exp-prime}
 \Psi= C_+\Phi^++C_2\Psi^++C_-\Phi^-+0\times\Psi^-.
\end{equation}
This obligatory zero coefficient in front of $\Psi^-$ or missing forth antisymmetric Bell state in the expansion (\ref{Bell-exp}) is related to restrictions imposed by the symmetry requirements for two-bozon states. Thus, even existence of biphoton qutrits as superpositions of only three basis wave functions occurs exclusively owing to the symmetry restrictions eliminating the forth (antisymmetric) basis Bell state.

\section{Density matrices}

The first step for finding the degree of entanglement is related to a transition from the wave function $\Psi$ to the density matrix of the same pure biphoton state $\rho=\Psi\Psi^\dag$. The full density matrix of the qutrit (\ref{qutrit-general}) can be presented in following two forms

\begin{gather}
 \label{rho}
 \setlength{\extrarowheight}{0.3cm}
 \begin{matrix}
 \rho=|C_1|^2\left({1\quad 0\atop{0\quad 0}}\right)_1\otimes\left({1\quad 0\atop{0\quad 0}}\right)_2
 +|C_3|^2\left({0\quad 0\atop{0\quad 1}}\right)_1\otimes\left({0\quad 0\atop{0\quad 1}}\right)_2\\
  +\frac{|C_2|^2}{2}\left[\left({1\quad 0\atop{0\quad 0}}\right)_1\otimes\left({0\quad 0\atop{0\quad 1}}\right)_2
 +\left({0\quad 0\atop{0\quad 1}}\right)_1\otimes\left({1\quad 0\atop{0\quad 0}}\right)_2\right.\\
 +\left.\left({0\quad 1\atop{0\quad 0}}\right)_1\otimes\left({0\quad 0\atop{1\quad 0}}\right)_2
 +\left({0\quad 0\atop{1\quad 0}}\right)_1\otimes\left({0\quad 1\atop{0\quad 0}}\right)_2\right]\\
 + C_1C_3^*\left({0\quad 1\atop{0\quad 0}}\right)_1\otimes\left({0\quad 1\atop{0\quad 0}}\right)_2+
  C_1^*C_3\left({0\quad 0\atop{1\quad 0}}\right)_1\otimes\left({0\quad 0\atop{1\quad 0}}\right)_2\\
 +\frac{C_1C_2^*}{\sqrt{2}}\left[\left({1\quad 0\atop{0\quad 0}}\right)_1\otimes\left({0\quad 1\atop{0\quad 0}}\right)_2+\left({0\quad 1\atop {0\quad 0}}\right)_1\otimes\left({1\quad 0\atop {0\quad 0}}\right)_2\right]\\
 +\frac{C_1^*C_2}{\sqrt{2}}\left[\left({1\quad 0\atop{0\quad 0}}\right)_1\otimes\left({0\quad 0\atop{1\quad 0}}\right)_2+\left({0\quad 0\atop {1\quad 0}}\right)_1\otimes\left({1\quad 0\atop {0\quad 0}}\right)_2\right]\\
 +\frac{C_3C_2^*}{\sqrt{2}}\left[\left({0\quad 0\atop{1\quad 0}}\right)_1\otimes\left({0\quad 0\atop{0\quad 1}}\right)_2+\left({0\quad 0\atop {0\quad 1}}\right)_1\otimes\left({0\quad 0\atop {1\quad 0}}\right)_2\right]\\
 +\frac{C_3^*C_2}{\sqrt{2}}\left[\left({0\quad 1\atop{0\quad 0}}\right)_1\otimes\left({0\quad 0\atop{0\quad 1}}\right)_2+\left({0\quad 0\atop {0\quad 1}}\right)_1\otimes\left({0\quad 1\atop {0\quad 0}}\right)_2\right]
 \end{matrix}
\end{gather}
 and
\begin{gather}
 \label{rho-4x4}
 \setlength{\extrarowheight}{0.3cm}
 \rho=\small{
 \left(
 \begin{matrix}
  |C_1|^2 & \frac{1}{\sqrt{2}}\,C_1C_2^* & \frac{1}{\sqrt{2}}\,C_1C_2^* & C_1C_3^*\\
  \frac{1}{\sqrt{2}}\,C_1^*C_2 & \frac{1}{2}\,|C_2|^2 & \frac{1}{2}\,|C_2|^2 & \frac{1}{\sqrt{2}}\,C_3^*C_2\\
  \frac{1}{\sqrt{2}}\,C_1^*C_2 & \frac{1}{2}\,|C_2|^2 & \frac{1}{2}\,|C_2|^2 & \frac{1}{\sqrt{2}}\,C_3^*C_2\\
  C_1^*C_3 & \frac{1}{\sqrt{2}}\,C_3C_2^* & \frac{1}{\sqrt{2}}\,C_3C_2^* & |C_3|^2
  \end{matrix}
  \right). }
\end{gather}
The next step is the reduction of the density matrix with respect to one of the photon variables, e.g., of the photon 2. Mathematically this means taking traces of all matrices with the subscript 2 in Eq. (\ref{rho}), which gives
\begin{equation}
  \rho_r=Tr_{_2}\rho=
  \label{reduced-qutr}
  \left(
 \begin{matrix}
  |C_1|^2+\displaystyle\frac{|C_2|^2}{2} &  \displaystyle\frac{C_1C_2^*+C_2C_3^*}{\sqrt{2}} \\
 \displaystyle\frac{C_1^*C_2+C_2^*C_3}{\sqrt{2}} & |C_3|^2+\displaystyle\frac{|C_2|^2}{2}
  \end{matrix}\right).
\end{equation}

It may be interesting to analyze a relation between the $4\times 4$ density matrix (\ref{rho-4x4}) and the $3\times 3$ coherence matrix introduced by Klyshko in 1997 \cite{Klyshko-97}. The density matrix $\rho$  (\ref{rho-4x4}) is written in a natural two-photon basis
\begin{equation}
 \label{natuaral basis}
 \left({1\atop 0}\atop{0\atop 0}\right),\;\left({0\atop 1}\atop{0\atop 0}\right),\;\left({0\atop 0}\atop{1\atop 0}\right),\;\left({0\atop 0}\atop{0\atop 1}\right).
\end{equation}
The question is how it can be transformed to the basis of states $\Psi_{HH}$, $\Psi_{VV}$, $\Psi_{HV}$ plus the empty antisymmetric states $\Phi^-$ (\ref{Bell-fourth})? Evidently, the transformation
\begin{gather}
 \nonumber
  \left({1\atop 0}\atop{0\atop 0}\right),\left({0\atop 1}\atop{0\atop 0}\right),\left({0\atop 0}\atop{1\atop 0}\right),\left({0\atop 0}\atop{0\atop 1}\right)\rightarrow\\
  \label{basis transf}
  \rightarrow\left({1\atop 0}\atop{0\atop 0}\right),\frac{1}{\sqrt{2}}\left({0\atop 1}\atop{1\atop 0}\right),\frac{1}{\sqrt{2}}\left({0\atop 1}\atop{-1\atop 0}\right),\left({0\atop 0}\atop{0\atop 1}\right)
\end{gather}
is provided by the matrix
\begin{equation}
 \label{matrix U}
 U=\left(
 \setlength{\extrarowheight}{0.1cm}
 \begin{matrix}
 1 & 0 & 0 & 0\\
 0 & \frac{1}{\sqrt{2}} & \frac{1}{\sqrt{2}} & 0\\
 0 & \frac{1}{\sqrt{2}} & -\frac{1}{\sqrt{2}} & 0\\
 0 & 0 & 0 & 1
 \end{matrix}
 \right).
\end{equation}
Now, transformed to the basis $\left\{\Psi_{HH}, \Psi_{VV}, \Psi_{HV}, \Psi^-\right\}$, the density matrix $\rho$ (\ref{rho-4x4})  takes the form
 \begin{equation}
 \label{rho-transf}
 \rho_{transf}=U\rho U=\left(
 \setlength{\extrarowheight}{0.1cm}
 \begin{matrix}
  |C_1|^2 & C_1C_2^* & 0 & C_1C_3^*\\
  C_1^*C_3 & |C_2|^2 & 0 & C_3^*C_2\\
  0 & 0 & 0 & 0\\
  C_1^*C_3 & C_3C_2^* & 0 & |C_3|^2
  \end{matrix}
  \right).
\end{equation}
A part of this matrix with nonzero rows and columns coincides with the $3\times 3$ coherence matrix \cite {Klyshko-97,Klyshko-99}
\begin{equation}
 \label{coh-matr}
 \rho_{coh}=\left(
 \setlength{\extrarowheight}{0.1cm}
 \begin{matrix}
  |C_1|^2 & C_1C_2^* & C_1C_3^*\\
  C_1^*C_3 & |C_2|^2 & C_3^*C_2\\
  C_1^*C_3 & C_3C_2^* & |C_3|^2
  \end{matrix}
  \right).
\end{equation}
Though the coherence matrix $\rho_{coh}$ (\ref{coh-matr}) is widely used and analyzed  in literature, both $\rho_{coh}$ and $\rho_{transf}$ are hardly appropriate for reduction over one of the photon variables (e.g., 2) and for finding correctly the reduced density matrix $\rho_r$ (\ref{reduced-qutr}) because the variables 1 and 2 are mixed not only in the matrix $\rho_{transf}$ itself but also in the transformed basis of Eq. (\ref{basis transf}).

\section{Degree of entanglement}

As known \cite{Grobe, Knight}, the trace of the squared reduced density matrix $\rho_r^2$ determines purity of the reduced state coinciding with the inverse value of the Schmidt entanglement parameter $K^{-1}$. The result of its calculation for the reduced density matrix of   Eq. (\ref{reduced-qutr}) is given by
\begin{gather}
  \nonumber
  K^{-1}=Tr(\rho_r^2)=\left(|C_1|^2+\frac{|C_2|^2}{2}\right)^2+\left(|C_3|^2+\frac{|C_2|^2}{2}\right)^2\\
  \label{K-qutrit}
  +|C_1^*C_2+C_2^*C_3|^2.
\end{gather}
With the normalization condition (\ref{norm-qtr}) taken into account, Eq. (\ref{K-qutrit}) can be reduced to a much simpler form
\begin{equation}
 \label{K-2}
  K=\frac{2}{2-|2C_1C_3-C_2^2|^2}.
\end{equation}

It's known also \cite{Rungta}, that in the case of bipartite states with the dimensionality of the one-particle  Hilbert space $d=2$, there is a simple algebraic relation between the Wootters' concurrence $C$ \cite{Wootters} and the Schmidt entanglement parameter $K$, owing to which
\begin{equation}
 \label{C-qutrit}
 C=\sqrt{2\left(1-\frac{1}{K}\right)}=|2C_1C_3-C_2^2|.
\end{equation}
At last, in terms of the constants $C_\pm$ (\ref{C-pm}), Eqs. (\ref{K-2}) and (\ref{C-qutrit}) take the form
\begin{equation}
 \label{K-C-tilde}
 K=\frac{2}{2-|C_+^2-C_-^2-C_2^2|^2},\;C=|C_+^2-C_-^2-C_2^2|.
\end{equation}

Note that the expressions for the concurrence $C$ [Eq. (\ref{C-qutrit}) and the last formula of Eq. (\ref{K-C-tilde})] can be derived also directly from the original Wootters' definition \cite{Wootters}. Indeed, for a pure bipartite state with the dimensionality of the one-particle Hilbert space  $d=2$,  the concurrence is defined in Ref. \cite{Wootters} as
\begin{equation}
 \label{Conc-Woo}
  C=|\langle\Psi|\widetilde{\Psi}^*\rangle|^2,
\end{equation}
where $\widetilde{\Psi}$ is the function or state vector arising from $\Psi$ after the ``spin-flip$"$ operation
\begin{equation}
 \label{spin-flip}
 \widetilde{\Psi}=\left(\sigma_y\right)_1\left(\sigma_y\right)_2\Psi,
\end{equation}
and $\sigma_y$ is the Pauli matrix, $\sigma_y=\left({{0\;-i}\atop{i\;\;\;0}}\right)$. For qutrits,  $\Psi$ is given by Eqs. (\ref{qutrit-general}) or (\ref{Bell-exp}). The rules of the ``spin-flip$"$ transformation for one-photon wave functions are $\left(1\atop 0\right)\rightarrow -i\left(0\atop 1\right)$ and $\left(0\atop 1\right)\rightarrow i\left(1\atop 0\right)$. From here we easily find the spin-flip transforms of the qutrit basis wave functions (\ref{columns-HH})-(\ref{columns-HV})
\begin{gather}
 \nonumber
 {\widetilde\Psi}_{HH}=-\Psi_{VV},\,{\widetilde\Psi}_{VV}=-\Psi_{HH},\,{\widetilde\Psi}_{HV}=\Psi_{HV},\\
 {\widetilde\Psi}_{+}=-\Psi_{+},\,{\widetilde\Psi}_-=\Psi_-
 \label{sp-fl-basis}
\end{gather}
and of the general-form qutrit wave function (\ref{qutrit-general})
\begin{gather}
 \nonumber
 {\widetilde\Psi}= -C_1\Psi_{VV}+C_2\Psi_{HV}-C_3\Psi_{HH}\\
 \label{sp-fl-wf}
 =-C_+\Psi_++C_-\Psi_-+C_2\Psi_{HV}.
\end{gather}
Substitution of these expressions into Eq. (\ref{Conc-Woo}) gives
\begin{equation}
 \label{C-via-C-pm}
 C^2=\left|-2C_1^*C_3^*+C_2^{*\,^2}\right|=\left|C_-^{*\,^2}-C_+^{*\,^2}+C_2^{*\,^2}\right|,
\end{equation}
in a complete agreement with Eqs. (\ref{C-qutrit}) and (\ref{K-C-tilde}).

\begin{figure}[h]
\centering\includegraphics[width=7cm]{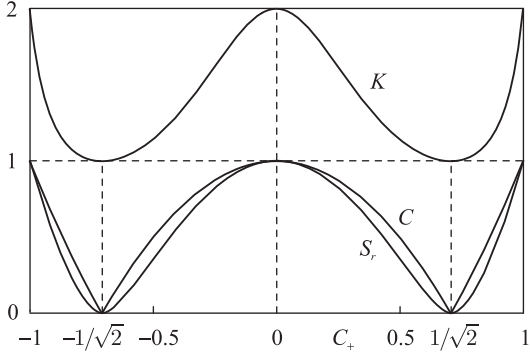}
\caption{{\protect\footnotesize {The Schmidt entanglement parameter $K_{real}$ (\ref{K-2}), concurrence $C_{real}$ (\ref{C-qutrit}) and the von Neumann subsystem entropy  $S_{r\,real}$ (\ref{entropy}) of the qutrit (\ref{qutrit-general}), (\ref{Bell-exp}) with real constants $C_{1,2,3}$, $C_\pm$ vs. $C_+$ (\ref{C-pm}).}}}\label{Fig1}
\end{figure}

In a special case of real constants $C_{1,2,3}$ and $C_\pm$, owing to normalization (\ref{norm-qtr}), the Schmidt entanglement parameter  and concurrence (\ref{K-C-tilde}) appear to be determined by the only real parameter  $C_+$
\begin{equation}
 \label{K-C-via C1}
 K_{real}=\frac{2}{1+4C_+^2-4C_+^4},\;C_{real}=|2C_+^2-1|.
\end{equation}
The functions $K_{real}(C_+)$ and $C_{real}(C_+)$ are shown in Fig. \ref{Fig1} together with the subsystem entropy found in the following section.
 This picture demonstrates clearly that qutrits are non-entangled ($K_{real}=1$ and $C_{real}=0$) if  $C_+=\pm 1/\sqrt{2}$ and, hence, $C_-^2+C_{HV}^2=\frac{1}{2}$. Consequently, the family of non-entangled qutrit wave functions with real coefficients is given by
\begin{gather}
  \nonumber
  \Psi_{NE\,real}=\frac{1}{\sqrt{2}}\Big(\Psi_+ +\sin\phi\,\Psi_{HV}+\cos\phi\,\Psi_-\Big)\\
  \label{Non-ent-wf-real}
  =\cos^2\left(\frac{\phi}{2}\right)\Psi_{HH}+\frac{\sin\phi}{\sqrt{2}}\Psi_{HV}
  +\sin^2\left(\frac{\phi}{2}\right)\Psi_{VV}
\end{gather}
with arbitrary $\phi$. If the constants $C_{1,2,3}$, $C_\pm$ are complex, a general condition of no-entanglement is the same: $C=0$. As seen from the general expression for $C$ (\ref{C-qutrit}), the concurrence depends on phases $\varphi_{1,2,3}$ of the constants $C_{1,2,3}$ only via the combination $\varphi_1+\varphi_3-2\varphi_2$. Hence, if $\varphi_2=\frac{1}{2}(\varphi_1+\varphi_3)$, in the case of complex constants $C_{1,2,3}$, Eq. (\ref{C-qutrit}) is reduced to $C=\Big|2|C_1||C_3|-|C_2|^2\Big|$, i.e. this case appears to be equivalent to the case of real constants $C_{1,2,3}$. From here we find that the general three-parametric family of wave functions of non-entangled qutrits is given by
\begin{gather}
  \nonumber
  \Psi_{NE\,general}(\phi,\varphi_1,\varphi_3)=\frac{\sin\phi}{\sqrt{2}}\,e^{\frac{i}{2}(\varphi_1+\varphi_3)}\Psi_{HV}\\
  \label{Non-ent-wf}
  +\cos^2\left(\frac{\phi}{2}\right)e^{i\varphi_1}\Psi_{HH}
  +\sin^2\left(\frac{\phi}{2}\right)e^{i\varphi_3}\Psi_{VV}
\end{gather}
with arbitrary $\phi$, $\varphi_1$, and $\varphi_3$. With these wave functions found explicitly we can reconstruct the corresponding family of state vectors of non-entangled qutrits
\begin{gather}
  \nonumber
  |\Psi\rangle_{NE\,general}=\frac{1}{\sqrt{2}}\Big\{\sin\phi\,e^{\frac{i}{2}(\varphi_1+\varphi_3)}a_H^\dag a_V^\dag\\
  \label{Non-ent-sv}
  +\cos^2\left(\frac{\phi}{2}\right)e^{i\varphi_1}a_H^{\dag^2}
  +\sin^2\left(\frac{\phi}{2}\right)e^{i\varphi_3}a_V^{\dag^2}\Big\}\,|0\rangle.
\end{gather}

Qutrits are maximally entangled when $C=1$ and $K=2$, and for the wave functions with real constants $C_{1,2,3}$, $C_\pm$ this occurs in two cases: $C_+=\pm 1$ and $C_+=0$. In the first of these cases $\Psi=\pm\Psi_+$. In the second case the maximally entangled wave function has a form of an arbitrary superposition of $\Psi_{HV}$ and $\Psi_-$, $\Psi=\sin\phi\,\Psi_{HV}+\cos\phi\,\Psi_-$. As previously, this result can be generalized for the case of wave functions with complex coefficients such that $\varphi_2=\frac{1}{2}(\varphi_1+\varphi_3)$. As a result we get the  following three-parametric family of wave functions of maximally entangled qutrits
\begin{gather}
 \nonumber
 \Psi_{\max}(\phi,\varphi_1,\varphi_3)=\sin\phi\,e^{\frac{i}{2}(\varphi_1+\varphi_3)}\Psi_{HV}\\
 \label{max-ent}
 +\frac{\cos\phi}{\sqrt{2}}\left(e^{i\varphi_1}\Psi_{HH}-e^{i\varphi_3}\Psi_{VV}\right)
\end{gather}
and the corresponding family of state vectors
\begin{gather}
 \nonumber
 |\Psi\rangle_{\max}=\Big\{\sin\phi\,e^{\frac{i}{2}(\varphi_1+\varphi_3)}a_H^\dag a_V^\dag\\
 \label{max-ent-sv}
 +\frac{\cos\phi}{2}\left(e^{i\varphi_1}a_H^{\dag^2}-e^{i\varphi_3}a_V^{\dag^2}\right)\Big\}\,|0\rangle.
\end{gather}
Eqs. (\ref{max-ent}) and (\ref{max-ent-sv}) are more general than the expression for the maximally entangled state of Ref. \cite{Genovese-Bell}, which follows from (\ref{max-ent-sv}) at $\phi=0$, $|\Psi_{\max}\rangle|_{\phi=0}=\frac{1}{\sqrt{2}}(e^{i\varphi_1}|2_H\rangle-e^{i\varphi_3}|2_V\rangle)$.

In the case $\phi=\pi/2$ Eqs. (\ref{max-ent}) and (\ref{max-ent-sv}) show that the state $\Psi_{HV}$ ($|1_H,1_V\rangle$) belongs to the family of maximally entangled states too (contrary to a met opinion that the state $|1_H,1_V\rangle$ is factorable). A way of seeing explicitly whether qutrits are factorable or not consists in finding their Schmidt decompositions, which can contain either two  products of Schmidt modes or only one product. This analysis is carried out in the following section.

But before switching to the Schmidt-mode analysis let us discuss briefly the problem of qutrit polarization. If we define the biphoton polarization vector, as suggested by Wang \cite{Wang}, ${\vec \xi}=Tr\left(\rho_r{\vec\sigma}\right)$, where ${\vec\sigma}$ is the vector of Pauli matrices, from Eq. (\ref{reduced-qutr}) we easily find
\begin{gather}
 \nonumber
 {\vec\xi}=\Big\{\sqrt{2}\,{\rm Re}(C_1C_2^*+C_2C_3^*),\,-\sqrt{2}\,{\rm Im}(C_1C_2^*+C_2C_3^*),\\
 \label{polar-vec}
 |C_1|^2-|C_3|^2\Big\}.
\end{gather}
A direct comparison with the results obtained in 1999 by Burlakov and Klyshko \cite{Klyshko-99} for polarization characteristics of qutrits shows that the polarization vector ${\vec\xi}$ (\ref{polar-vec}) coincides exactly with one half of the vector of Stokes parameters ${\vec S}=\{S_1,\,S_2,\,S_3\}$ of Ref. \cite{Klyshko-99}, and the absolute value of ${\vec\xi}$ coincides with the degree of polarization $P$ (introduced also in \cite{Klyshko-99})
\begin{equation}
 \label{Wang-Klyshko}
 {\vec\xi}=\frac{1}{2}\,{\vec S},\quad \left|{\vec\xi}\,\right|=P=\frac{1}{2}\sqrt{S_1^2+S_2^2+S_3^2}.
\end{equation}
Note that the Stokes parameter and the degree of polarization were found in Ref. \cite{Klyshko-99} in a way, absolutely different from that used above for derivation of the polarization vector ${\vec\xi}$. For this reason the coincidences (\ref{Wang-Klyshko}) are rather non-trivial.

The derived expression for the polarization vector ${\vec\xi}$ (\ref{polar-vec}) can be compared with the general expression for the qutrit concurrence $C$ (\ref{C-qutrit}) to show that they obey the relation ${\vec\xi}\,^{\textstyle\,^2}=1-C^2$ obtained by Wang in 2000 \cite{Wang}. Note, however, that for qutrits we have  used the concurrence of Eq. (\ref{C-qutrit}), $C=|2C_1C_3-C_2^2|$, rather than $2|ad-bc|$ obtained in Refs. \cite{Wang} for the state $|\Psi\rangle_{abcd}=a|00\rangle+b|01\rangle+c|10\rangle+d|11\rangle$ to be commented below in section 7. In terms of the degree of polarization $P$ the relation ${\vec\xi}\,^{\textstyle\,^2}=1-C^2$ can be rewritten as
\begin{equation}
 \label{pol-ent}
 C^2+P^2=1.
\end{equation}
Thus, the degrees of polarization and entanglement anti-correlate with each other: the maximally entangled qutrits are non-polarized, and maximally polarized states are non-entangled. Though, maybe, intuitively more or less expected, as far as we know, anticorrelation of polarization and entanglement has never been presented in a rigorous mathematical form of Eq. (\ref{pol-ent}).

\section{Schmidt modes of qutrits and the subsystem entropy}

Entanglement means that the biphoton wave function cannot be factorized whereas no-entanglement means that it is factorable. A transition from non-factorable to factorable wave functions can be reasonably explained in terms of Schmidt modes. Schmidt modes are eigenfunctions of the reduced density matrix, i.e., solutions of the equation $\rho_r\psi=\lambda\psi$. As in the case of qutrits $\rho_r$ (\ref{reduced-qutr}) is the $2\times 2$ matrix, it has two eigenvalues $\lambda_\pm$, its eigenfunctions are 2-row columns $\psi_\pm=\left({a_\pm\atop b_\pm}\right)$, and the eigenvalue-eigenfunction equation has the form
\begin{equation}
 \label{eigen-eq}
 \rho_r\left({a_\pm\atop b_\pm}\right)=\lambda_\pm\left({a_\pm\atop b_\pm}\right).
\end{equation}
The modes can be normalized $|a_\pm|^2+|b_\pm|^2=1$ and they are orthogonal to each other $a_+a_-^*+b_+b_-^*=0$. In terms of $\lambda_\pm$ the Schmidt entanglement parameter equals to $K=\left(\lambda_+^2+\lambda_-^2\right)^{-1}$.

In accordance with the Schmidt theorem, the biphoton wave function can be presented as a sum of products of Schmidt modes (Schmidt decomposition). In the case $d=2$ (two-qubit states or qutrits) the Schmidt decomposition contains only two terms
\begin{equation}
 \label{Schmidt theorem}
 \Psi=\sum_\pm\sqrt{\lambda_\pm}\,\psi_{\pm}(1)\psi_{\pm}(2),
\end{equation}
where arguments of the Schmidt modes indicate variables of photons ``1$"$ and ``2$"$.
This decomposition shows that in a general case the wave function $\Psi$ is nonseparable. Exceptions occur  when one of the eigenvalues of the reduced density matrix, $\lambda_+$ or $\lambda_-$, becomes equal zero.

Eigenvalues of the matrix (\ref{reduced-qutr}) can be found rather easily and they can be reduced to a very simple form being expressed via the concurrence $C$
\begin{equation}
 \label{eigenvalues}
 \lambda_\pm=\frac{1}{2}\left(1\pm\sqrt{1-C^2}\right).
\end{equation}

In the case $C=1$, when entanglement is maximal, $\lambda_+=\lambda_-=\frac{1}{2}$, i.e., two products in the Schmidt decomposition (\ref{Schmidt theorem}) are presented with equal weights and, clearly, the wave function is nonseparable.

In the case $C=0$ (no entanglement) Eq. (\ref{eigenvalues}) gives $\lambda_+=1$ and $\lambda_-=0$. In this case one of the products of Schmidt modes ($\psi_-\psi_-$) in the Schmidt decomposition disappears because the coefficient $\sqrt{\lambda_-}$ $\,$in front of it vanishes. This is the reason of separability of the wave function in the case of no-entanglement. As for the remaining product in the Schmidt decomposition, the corresponding eigenfunction of the reduced density matrix $\psi_+$  at $C=0$ can be  found easily and has a reasonably simple form
(\ref{Non-ent-wf})
\begin{equation}
 \label{Schmidt-mode-C=0}
 \psi_+=\left({a_+\atop b_+}\right)=\left({\cos(\phi/2)e^{\frac{i}{2}\varphi_1}\atop \sin(\phi/2)e^{\frac{i}{2}\varphi_3}}\right).
\end{equation}
With this expression we find immediately that in the case $C=0$ the Schmidt decomposition (\ref{Schmidt theorem}) gives the following representation for the factored biphoton wave function
\begin{equation}
 \label{factorization}
 \Psi_{NE}=\left({\cos(\phi/2)e^{\frac{i}{2}\varphi_1}\atop \sin(\phi/2)e^{\frac{i}{2}\varphi_3}}\right)_1\left({\cos(\phi/2)e^{\frac{i}{2}\varphi_1}\atop \sin(\phi/2)e^{\frac{i}{2}\varphi_3}}\right)_2.
\end{equation}
Identity of Eqs. (\ref{Non-ent-wf}) and (\ref{factorization}) can be checked directly, be means of substitution in Eq. (\ref{Non-ent-wf}) instead of $\Psi_{HH}$, $\Psi_{HV}$, $\Psi_{VV}$ their column representations (\ref{columns-HH})-(\ref{columns-HV}), calculation of direct products of columns in all terms of Eq. (\ref{Non-ent-wf}) and in Eq. (\ref{factorization}), and presenting $\Psi_{NE}$ in both cases in the form of the following 4-row column
\begin{equation}
 \label{wf-ne-4rows}
 \Psi_{NE\,general}(\phi,\varphi_1,\varphi_3)=\left(
 \begin{matrix}
 \cos^2(\phi/2)e^{i\varphi_1}\\
 \frac{1}{2}\sin(\phi)e^{i(\varphi_1+\varphi_3)}\\
 \frac{1}{2}\sin(\phi)e^{i(\varphi_1+\varphi_3)}\\
 \sin^2(\phi/2)e^{i\varphi_3}
 \end{matrix}
 \right).
\end{equation}

Eq. (\ref{eigenvalues}) can be used to find the subsystem entropy \cite{Barnett} defined as
\begin{equation}
 \label{entropy}
 S_r=-Tr\left(\rho_r\log_2\rho_r\right)=-\sum_\pm\lambda_\pm\log_2\lambda_\pm.
\end{equation}
In the case of qutrits with real coefficients the concurrence $C_{real}$ itself is a function of $C_+$ (\ref{K-C-via C1}). With $C_{real}(C_+)$ substituted instead of $C$ into Eq. (\ref{eigenvalues}) and then $\lambda_\pm[C(C_+)]$ substituted into Eq. (\ref{entropy}), we get a function $S_{r\, real}(C_+)$, which is plotted in Fig. \ref{Fig1}. Though different from both $C(C_+)$ and $K(C_+)$, the subsystem entropy $S_{r\,real}(C_+)$ has the same main features as two other entanglement quantifiers: $S_{r\,real}$ is minimal and equals zero at $C_+=\pm 1/\sqrt{2}$ and $S_{r\,real}$ is maximal and equals 1 at $C_+=0$ and $C_+=\pm 1$. Therefore, though the Schmidt entanglement parameter $K$, concurrence $C$ and the subsystem entropy $S_r$ characterize the degree of entanglement in different metrics, their behavior is very similar, which confirms all conclusions made above about conditions of separability and nonseparability of qutrits and their wave functions.

As mentioned above, one of the basis wave functions of qutrits, $\Psi_{HV}$ (\ref{HV}), is maximally entangled ($K=2$, $C=1$) and, hence, unseparable. Eigenvalues of the reduced density matrix of this state are degenerate, $\lambda_+=\lambda_-=\frac{1}{2}$, and the Schmidt decomposition has the form
\begin{equation}
 \label{HV-Schm-decomp}
 \Psi_{HV}=\frac{\psi_+(1)\psi_+(2)+\psi_-(1)\psi_-(2)}{\sqrt{2}}
\end{equation}
with the Schmidt modes given by
\begin{equation}
 \label{HV-Schm-modes}
 \psi_+^{HV}=\frac{1}{\sqrt{2}}\left({1\atop 1}\right),\;\psi_-^{HV}=\frac{1}{\sqrt{2}}\left({1\atop -1}\right).
\end{equation}

 Thus, entanglement of the state $\Psi_{HV}$ (\ref{HV}) is proved here in several ways, by direct calculations of the entanglement quantifiers $K$, $C$ and $S_r$, and by showing that its Schmidt decomposition contains two products of the Schmidt modes (\ref{HV-Schm-decomp}). In literature the opinion about entanglement of the state $\Psi_{HV}$ is shared by some authors \cite{Lanyon, Paskauskas}, though sometimes the state $\Psi_{HV}$ is claimed to be separable \cite{Li}. Actually, this difference of opinions reflects a rather popular point of view that all basis states of qutrits and ququarts (see below) are separable and non-entangled. In a more general form, this statements can be reformulated as saying that all states of the type $a_i^\dag a_j^\dag|0\rangle$ generate separable wave functions in both cases of coinciding $i=j$ and non-coinciding $i\neq j$ mode indices. Our analysis shows that for biphoton states the latter is not true. In reality, for arbitrary photon variables (polarization, angular, frequency, or their combinations), if $\psi_i$ and $\psi_j$ are the wave functions of the $i-th$ and $j-th$ modes and $i\neq j$, the state vector $a_i^\dag a_j^\dag|0\rangle$  generates the wave function of the type $\Psi_{HV}$ (\ref{HV}),
\begin{equation}
 \label{psi-ij-genertal}
 \Psi_{ij}=\frac{\psi_i(1)\psi_j(2)+\psi_i(2)\psi_j(1)}{\sqrt{2}},
\end{equation}
which is entangled with the degree of entanglement characterized by $K=2$, $C=S_r=1$, and the Schmidt decomposition of the type (\ref{HV-Schm-decomp}), (\ref{HV-Schm-modes}).

\section{Finding the degree of entanglement of qutrits from direct polarization measurements}

\begin{figure}[h]
\centering\includegraphics[width=8cm]{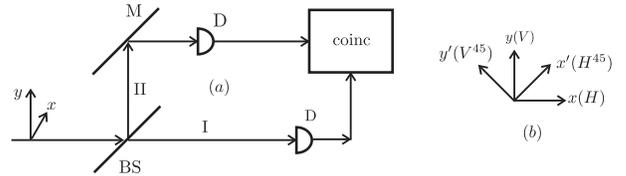}
\caption{{\protect\footnotesize {$(a)$ A scheme of and $(b)$ two bases for coincidence measurements; BS denotes beam splitter, M - mirror and D - detectors.}}}\label{Fig2}
\end{figure}

In this section we will show that there is a rather simple method of measuring experimentally the degree of entanglement of qutrits (supposedly not known in advance). The key idea consists in splitting the original biphoton beam for two identical parts by a non-selective beam splitter (BS) and performing a series of coincidence photon-counting measurements in the arising channels I an II (see Fig. \ref{Fig2}$a$).
Measurements have to be done with polarizers installed in the horizontal ($x$) and vertical ($y$) directions, as well as along the axes $x^\prime$ and $y^\prime$ turned for the angle $45^\circ$ with respect to, correspondingly, $x$- and $y$-axes (Fig. \ref{Fig2}$b$)). As shown below this set of measurements is sufficient for determining the degree of entanglement as well as for a full reconstruction of all parameters of arbitrary biphoton qutrits.

\subsection{Beam splitter}

In terms of biphoton wave functions, BS adds an additional degree of freedom to each photon - the propagation angles $\theta_1$ and $\theta_2$, which can take only two value, $0$ and $90^\circ$. In a nonselective BS each photon has equal probabilities of transmitting or being reflected. Thus, if the biphoton wave function in front of BS is $\Psi$, after BS it takes the form
\begin{gather}
 \nonumber
 \Psi_{\rm after BS}=\Psi\times\textstyle\frac{1}{2}(\delta_{\theta_1,\,0}
 -\delta_{\theta_1,\,90^\circ})(\delta_{\theta_2,\,0}-\delta_{\theta_2,\,90^\circ})\\
 \label{bs}
 \setlength{\extrarowheight}{0.2cm}
  \begin{matrix}
 \equiv\Psi\otimes\frac{1}{2}\left({1\atop -1}\right)_1^{(\theta)}
 \otimes\left({1\atop -1}\right)^{(\theta)}_2,
 \end{matrix}
\end{gather}
As the angular variables in $\Psi_{\rm after BS}$ are separated from polarization variables of $\Psi$ and, besides, parts depending on the variables $\theta_1$ and $\theta_2$ are also factorized, the angular factor in $\Psi_{\rm after BS}$ (\ref{bs}) does not add any additional entanglement to this wave function compared to $\Psi$. I.e., the nonselective BS itself does not increase or diminish the degree of entanglement of any biphoton states. Coincidence measurements in the state arising after BS are suggested here as a tool for determining qutrit parameters in front of BS and the  degree of entanglement of the original biphoton state.

As it follows from Eq. (\ref{bs}) and from  the described features of BS, the latter splits between the channels I and II only photons of a half of all incoming biphoton pairs, whereas another half of pairs remains unsplit, and these unsplit pairs are equally distributed between the channels I and II. Let $N^{tot}_{HH}$, $N^{tot}_{VV}$, and $N^{tot}_{HV}$ be large ($\gg1$) total amounts of biphoton pairs with coinciding ($HH$, $VV$) and differing ($HV$) polarizations of photons  generated in a crystal per some given time. Then the total amount of generated pairs is  $N^{tot}_{pairs}=N^{tot}_{HH}+N^{tot}_{VV}+N^{tot}_{HV}$ and the total amount of generated photons is $N^{tot}_{phot}=2N^{tot}_{pairs}$. Among all these generated photons there are $N^{tot}_{H}=2N^{tot}_{HH}+N^{tot}_{HV}$ horizontally and $N^{tot}_{V}=N^{tot}_{HV}+2N^{tot}_{VV}$ vertically polarized photons. Relative amounts of horizontally and vertically polarized photons can be interpreted as single-particle (absolute) probabilities for photons to have horizontal and vertical polarizations
\begin{gather}
 \nonumber
 w^{(s)}_H=\frac{N^{tot}_{H}}{N^{tot}_{phot}}=\frac{N^{tot}_{HH}+\frac{1}{2}N^{tot}_{HV}}{N^{tot}_{HH}
 +N^{tot}_{VV}+N^{tot}_{HV}}\\
 \label{rel-prob-via-N-tot}
 w^{(s)}_V=\frac{N^{tot}_{V}}{N^{tot}_{phot}}=\frac{N^{tot}_{VV}+\frac{1}{2}N^{tot}_{HV}}{N^{tot}_{HH}
 +N^{tot}_{VV}+N^{tot}_{HV}}.
\end{gather}
Though these equations are nice, unfortunately, in experiment amounts of generated pairs are not directly observable.

\begin{figure}[h]
\centering\includegraphics[width=8cm]{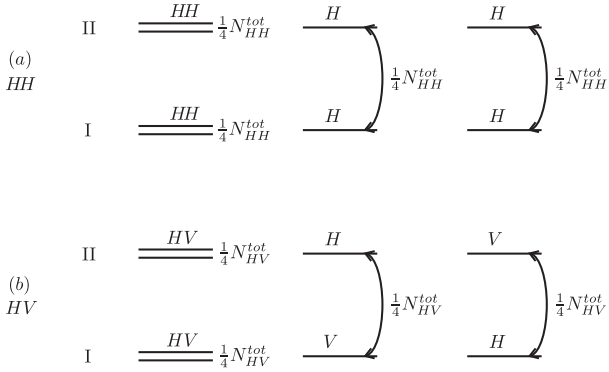}
\caption{{\protect\footnotesize {Distribution of photons in channels I and II after BS for $(a)\,HH$} and $(b)\,HV$ pairs; double lines are unsplit pairs, arcs with arrows indicate photons of split pairs.}}\label{Fig3}
\end{figure}

After BS we get $\frac{1}{4}N^{tot}_{\sigma\sigma^\prime}$ unsplit pairs in each channel, and photons of the remaining $\frac{1}{2}N^{tot}_{\sigma\sigma^\prime}$ split pairs are equally distributed between the channels I and II (Fig. \ref{Fig3}).
Owing to unsplit pairs results of photon counting by a single detector in one of the channels I or II can be rather confusing, because when an unsplit pair comes to the detector the latter can produce one click instead of two. In contrast, coincidence measurements by two detectors in channels I and II [Fig. \ref{Fig2}$a$] register only photons of split pairs, which makes such measurements quite unambiguous and informative.

Relations between amounts of single photons and the corresponding total amounts of pairs are different in the cases when single photons arise from split $HH$ and $VV$ pairs, and when they arise from split $HV$ pairs. As seen from Fig. \ref{Fig3}, the amount of $H$- or $V$-polarized photons arising from the split $HH$ or $VV$ pairs equals to $\frac{1}{2}N^{tot}_{HH}$ or $\frac{1}{2}N^{tot}_{VV}$ (Fig. \ref{Fig3}$a$), whereas amounts of the same photons arising from the split $HV$ pairs is equal to $\frac{1}{4}N^{tot}_{HV}$ (Fig. \ref{Fig3}$b$). From these results we easily find amounts of detector coincidence counts $\left.N^{d\,(c)}_\sigma\right|_{\sigma^\prime}$ obtainable when  $\sigma$- and $\sigma^\prime$-polarizers are installed in front of detectors, correspondingly, in the channels I and II
\begin{gather}
 \nonumber
 \left.N^{d\,(c)}_H\right|_{H}=\frac{\eta}{2}N^{tot}_{HH},\;
 \left.N^{d\,(c)}_V\right|_{V}=\frac{\eta}{2}N^{tot}_{VV},\\
 \label{counts-pairs}
 \left.N^{d\,(c)}_H\right|_{V}=\left.N^{d\,(c)}_V\right|_{H}=\frac{\eta}{4}N^{tot}_{HV},
\end{gather}
where $\eta$ is the efficiency of detectors.
Relative amounts of coincidence counts have sense of conditional probabilities $\left.w^{(c)}_\sigma\right|_{\sigma^\prime}$ for a photon ``1$"$ to have polarization $\sigma$ under the condition that another photon, ``2$"$, has polarization $\sigma^\prime$. They are defined as  ratios of specific coincidence detector counts  (\ref{counts-pairs}) to the total amounts of coincidence counts
\begin{equation}
 \label{cond-prob-am-of-pairs}
 \left.w^{(c)}_\sigma\right|_{\sigma^\prime}=\frac{\left.N^{d\,(c)}_\sigma\right|_{\sigma^\prime}}
 {\sum_{\sigma,\sigma^\prime}\left.N^{d\,(c)}_\sigma\right|_{\sigma^\prime}},
\end{equation}
where
\begin{equation}
 \label{total-coi}
 \sum_{\sigma,\sigma^\prime}\left.N^{d\,(c)}_\sigma\right|_{\sigma^\prime}
 =\frac{\eta}{2}\big[N_{HH}^{tot}+N_{HV}^{tot}+N_{VV}^{tot}\big]=\frac{\eta}{2}N_{pair}^{tot}.
\end{equation}
Evidently, the sum of all conditional probabilities (\ref{cond-prob-am-of-pairs}) equals unit
\begin{equation}
 \label{norm-cond-prob}
 \sum_{\sigma,\sigma^\prime}\left.w^{(c)}_\sigma\right|_{\sigma^\prime}=
 \left.w^{(c)}_H\right|_{H}+2\left.w^{(c)}_H\right|_{V}+\left.w^{(c)}_V\right|_{V}=1
\end{equation}

Eqs. (\ref{cond-prob-am-of-pairs}) and (\ref{total-coi}) show that, in principle, experimental measurement of conditional probabilities is straightforward because they are expressed explicitly in terms of the coincidence detector counts $\left.N^{d\,(c)}_\sigma\right|_{\sigma^\prime}$. In terms of amounts of pairs in a beam expressions for conditional probabilities take the form
\begin{gather}
 \nonumber
 \left.w_H^{(c)}\right|_{H}=
 \frac{N_{HH}^{tot}}{N_{pair}^{tot}},\; \left.w_V^{(c)}\right|_{V}=
 \frac{N_{VV}^{tot}}{N_{pair}^{tot}}, \\
 \label{prob-vs-counts}
 \left.w_H^{(c)}\right|_{V}=\left.w_V^{(c)}\right|_{H}=
 \frac{N_{VH}^{tot}}{2N_{pair}^{tot}}.
\end{gather}
By the definition of conditional probabilities, their sums can be constructed to give single-particle, unconditional probabilities:
\begin{gather}
 \nonumber
 w_H^{(s)}=\left.w_H^{(c)}\right|_{H}+\left.w_H^{(c)}\right|_{V}
 =\frac{N_{HH}^{tot}+\frac{1}{2}N_{HV}^{tot}}{N_{pair}^{tot}}\\
 \label{single-prob}
 w_V^{(s)}=\left.w_V^{(c)}\right|_{V}+\left.w_V^{(c)}\right|_{H}
 =\frac{N_{VV}^{tot}+\frac{1}{2}N_{HV}^{tot}}{N_{pair}^{tot}}.
\end{gather}
These expressions are absolutely identical to Eqs. (\ref{rel-prob-via-N-tot}) derived simply from counting amounts of photons in the original biphoton beam without any beam splitter. Coincidence of Eqs. (\ref{rel-prob-via-N-tot}) and (\ref{single-prob}) proves that, indeed, coincidence measurements after BS can be used for getting information about such features of the beam before BS as characterizing it unconditional (single-particle) probabilities for photons to have horizontal or vertical polarizations. In other words, Eqs. (\ref{single-prob}) show that though owing to unsplit pairs direct measurement of single particle probabilities after BS is problematic, nevertheless, the probabilities $w_H^{(s)}$ and $w_V^{(s)}$ can be found from results of coincidence measurements.

Conditional probabilities of Eq. (\ref{prob-vs-counts}) can be used for a partial reconstruction of the qutrit parameters from experimental data. Indeed, directly from Eq. (\ref{rho}) for the density matrix $\rho$  we find the following relations between the absolute values of all three coefficients $C_{1,2,3}$ in the wave function (\ref{qutrit-general}) and the conditional probabilities
\begin{gather}
 \setlength{\extrarowheight}{0.2cm}
 \begin{matrix}
 \left.w_H^{(c)}\right|_{H}=|C_1|^2,\; \left.w_V^{(c)}\right|_{V}=|C_3|^2\\
 \left.w_V^{(c)}\right|_{H}=\left.w_H^{(c)}\right|_{V}=\frac{1}{2}|C_2|^2.
 \end{matrix}
 \label{cond.prob}
\end{gather}
For single-particle probabilities (\ref{single-prob}) relations with parameters of qutrits $C_i$ have the form
\begin{equation}
\label{sing-viqa Ci.prob}
 w_H^{(s)}=|C_1|^2+\frac{1}{2}|C_2|^2,\;w_V^{(s)}=|C_3|^2+\frac{1}{2}|C_2|^2.
\end{equation}

\subsection{Qutrits in a turned basis}

In addition to absolute values of $C_i$, we have to find their phases, $\varphi_{1,2,3}$, and this requires additional measurements, e.g. with polarizers installed at $45^\circ$ to $x$- and $y$-axes, as shown in Fig. \ref{Fig2}$b$. In theory, this is equivalent to description of the biphoton wave function in a turned basis. In a general case, let $x^{\alpha}$ and $y^{\alpha}$ be axes turned for an angle $\alpha$ with respect to $x$- and $y$-axes. Let polarizations along the axes $x^{\alpha}$ and $y^{\alpha}$ be denoted as $H^{\alpha}$ and $V^{\alpha}$. The corresponding one-photon wave functions are $\left({1\atop 0}\right)^{(\alpha)}$ and $\left({0\atop 1}\right)^{(\alpha)}$. From these one-photon wave functions we can construct the two-photon basis wave functions $\Psi_{H^\alpha H^\alpha}$, $\Psi_{H^\alpha V^\alpha}$, and $\Psi_{V^\alpha V^\alpha}$ in the same way (\ref{columns-HH})-(\ref{columns-HV}) as $\Psi_{HH}$, $\Psi_{HV}$, and $\Psi_{VV}$ were constructed from the one-photon wave functions $\left({1\atop 0}\right)$ and $\left({0\atop 1}\right)$. Let us present the wave function (\ref{qutrit-general})  in terms of expansion in the basis wave function of the frame ($x^\alpha,y^\alpha$)
\begin{gather}
 \Psi=C_1(\alpha)\Psi_{H^\alpha H^\alpha}+C_2(\alpha)\Psi_{H^\alpha V^\alpha}+C_3(\alpha)\Psi_{V^\alpha V^\alpha}
 \label{qutrit-alpha}.
\end{gather}
One-photon wave functions in the $\alpha$- and original frames are related with each other by the evident transformation formulas
\begin{gather}
 \label{transformation}
 \begin{matrix}
 \left({1\atop 0}\right)=\cos\alpha\left({1\atop 0}\right)^{(\alpha)}-\sin\alpha\left({0\atop 1}\right)^{(\alpha)},\\
 \left({0\atop 1}\right)=\sin\alpha\left({1\atop 0}\right)^{(\alpha)}+\cos\alpha\left({0\atop 1}\right)^{(\alpha)}.
 \end{matrix}
\end{gather}
By applying these formulas to all terms and columns in Eq. (\ref{qutrit-general}) and regrouping the arising  $\alpha$-frame column products we reduce finally the biphoton wave function (\ref{qutrit-general}) to the form (\ref{qutrit-alpha}) with the following relations between the coefficients $C_{1,2,3}(\alpha)$ and $C_{1,2,3}$
\begin{gather}
 \nonumber
 C_1(\alpha)=\cos^2\alpha \,C_1+\sqrt{2}\cos\alpha\sin\alpha \,C_2+\sin^2\alpha \,C_3,\\
 \nonumber
 C_2(\alpha)=-\sqrt{2}\cos\alpha\sin\alpha\, (C_1-C_3)+\cos 2\alpha \,C_2,\\
 \label{C123-alpha}
 C_3(\alpha)=\sin^2\alpha \,C_1-\sqrt{2}\cos\alpha\sin\alpha\,C_2+\cos^2\alpha\, C_3.
\end{gather}
Note that similar transformation formulas for the constants $C_\pm$ (\ref{C-pm}) have the form
\begin{gather}
 \label{C+(alpha)}
 C_+(\alpha)=C_+,\\
 \label{C-(alpha)-C2}
 \left\{{C_-(\alpha)=\cos 2\alpha \,C_-+\sin 2\alpha \,C_2
 \atop{C_2(\alpha)=-\sin 2\alpha\, C_-+\cos 2\alpha \,C_2}}.\right.
\end{gather}
This means, in particular, that
\begin{gather}
 \nonumber
 \Psi_+^{(\alpha)}\equiv\Psi_+,\\
 \nonumber
 \Psi_-^{(\alpha)}= \cos 2\alpha\Psi_-+\sin 2\alpha\Psi_{HV},\\
 \label{alpha-transf-funcs}
 \Psi_{HV}^{(\alpha)}= \cos 2\alpha\Psi_{HV}-\sin 2\alpha\Psi_-,
\end{gather}
i.e. the function $\Psi_+^{(\alpha)}$ has the same form in all $\alpha$-frames, whereas the functions $\Psi_-^{(\alpha)}$ and  $\Psi_{HV}^{(\alpha)}$ transform into each other with changing $\alpha$.

In a special case $\alpha=45^{\circ}$ Eqs. (\ref{C123-alpha}) are reduced to
\begin{gather}
 C_{1,3}(45^{\circ})=\frac{C_1+C_3}{2}\pm\frac{C_2}{\sqrt{2}},\,
 \label{C123-45}
 C_2(45^{\circ})=\frac{C_1-C_3}{\sqrt{2}}.
\end{gather}

 Similarly to (\ref{cond.prob}), the qutrit parameters in the basis turned for $45^\circ$ can be expressed in terms of the corresponding conditional probabilities
 \begin{gather}
 \setlength{\extrarowheight}{0.2cm}
 \begin{matrix}
 \left.w_{H^{45}}^{(c)}\right|_{H^{45}}=|C_1(45^\circ)|^2,\; \left.w_{V^{45}}^{(c)}\right|_{V^{45}}=|C_3(45^\circ)|^2\\
 \left.w_{V^{45}}^{(c)}\right|_{H^{45}}=\left.w_{H^{45}}^{(c)}\right|_{V^{45}}=\frac{1}{2}\,|C_2(45^\circ)|^2.
  \end{matrix}
 \label{abs-values-45}
\end{gather}
In their turn, the conditional probabilities $\left.w_{\sigma^{45}}^{(c)}\right|_{\sigma^{\prime\,45}}$ are related by equations identical to (\ref{prob-vs-counts}) to the corresponding amounts of counts with polarizers installed in front of detectors along the directions of the $x^\prime$ and $y^\prime$ axes in Fig. \ref{Fig2}$b$
\begin{equation}
 \left.w_{\sigma^{45}}^{(c)}\right|_{\sigma^{\prime\,45}}=\frac{\left.N^{d\,(c)}_{\sigma^{45}}\right|_{\sigma^{\prime\,45}}}
 {\sum_{{\sigma^{45}},\sigma^{\prime\,45}}\left.N_{\sigma^{45}}^{d\,(c)}\right|_{\sigma^{\prime\,45}}}.
  \label{prob-vs-counts-45}
\end{equation}
Similarly to (\ref{single-prob}) and (\ref{sing-viqa Ci.prob}) we can define also single-particle probabilities via conditional ones in a turned basis
\begin{gather}
 \nonumber
 w_{H^{45}}^{(s)}=\left.w_{H^{45}}^{(c)}\right|_{H^{45}}+\left.w_{H^{45}}^{(c)}\right|_{V^{45}}=|C_1(45^\circ)|^2,\\
 \label{sing-45}
 w_{V^{45}}^{(s)}=\left.w_{V^{45}}^{(c)}\right|_{V^{45}}+\left.w_{V^{45}}^{(c)}\right|_{H^{45}}=|C_3(45^\circ)|^2.
\end{gather}

\subsection{Concurrence and full reconstruction of qutrit parameters from experimental data}

In terms of absolute values of the qutrit parameters $C_{1,2,3}$ and their phases $\varphi_{1,2,3}$, the squared concurrence $C$ Eq. (\ref{C-qutrit}) has the form
\begin{gather}
 \nonumber
 C^2=4|C_1|^2|C_3|^2+|C_2|^4\\
 \label{conc-squared}
 -4|C_1||C_3||C_2|^2\cos(\varphi_1+\varphi_3-2\varphi_2).
\end{gather}
As well known and mentioned above, the common phase of the biphoton wave function does not affect any measurements. This means that multiplication of $\Psi$ by an arbitrary phase factor $e^{i\varphi_0}$ does not change any physical results. Let us take, for example, $e^{i\varphi_0}=-\varphi_2$, which makes the parameter $e^{i\varphi_0}C_2$ real, with corresponding changes of phases in two other parameters, $C_1$ and $C_3$. Equivalently, keeping in mind this procedure we simply can take $C_2$ real and, in a general case, $C_{1,3}$ complex,| $C_1=|C_1|e^{i\varphi_1}$ and $C_3=|C_3|e^{i\varphi_3}$. With real $C_2$ ($\varphi_2=0$), Eq. (\ref{conc-squared}) takes the form
\begin{gather}
 \nonumber
 C^2=|2C_1C_3-C_2^2|^2\\
 \label{conc-squared-simpler}
 =4|C_1|^2|C_3|^2+C_2^4-4|C_1||C_3|C_2^2\cos(\varphi_1+\varphi_3).
\end{gather}
For finding phases $\varphi_1$ and $\varphi_3$ we can use Eqs. (\ref{C123-45}), the second of which gives
\begin{gather}
 \nonumber
 |C_2(45^{\circ})|^2=\frac{|C_1-C_3|^2}{2}\\
 \label{eq-for-phases}
 =\frac{|C_1|^2+|C_3|^2}{2}-|C_1||C_3|\cos(\varphi_1-\varphi_3).
\end{gather}
One equation for two unknown phases $\varphi_1$ and $\varphi_3$ can be obtained from the difference of squared absolute values of the parameters $C_1(45^\circ)$ and $C_3(45^\circ)$ in Eqs. (\ref{C123-45}):
\begin{gather}
 \nonumber
 |C_1(45^\circ)|^2-|C_3(45^\circ)|^2=\sqrt{2}\,C_2\,{\rm Re}(C_1+C_3)\\
 \label{Eq-for-phi2}
 =\sqrt{2}\,C_2\Big[|C_1|\cos\varphi_1+|C_3|\cos\varphi_3\Big].
\end{gather}
In a general case, Eqs. (\ref{eq-for-phases}) and (\ref{Eq-for-phi2}) hardly can be further simplified to yield a simple analytical formula for the squared concurrence (\ref{conc-squared}) in terms of the experimentally measurable conditional probabilities. But numerical solution of Eqs. (\ref{eq-for-phases}) and (\ref{Eq-for-phi2}) with all parameters $|C_i|$ and $|C_i(45^\circ)|$ known does not represent any problems. Thus, by getting a full set of measurements of coincidence counts in two bases ($xy$) and ($x^\prime y^\prime $) (Fig. 2$b$) one can find all conditional probabilities $\left.w_{\sigma}^{(c)}\right|_{\sigma^\prime}$ (\ref{prob-vs-counts}) and $\left.w_{\sigma^{45}}^{(c)}\right|_{\sigma^{\prime\,45}}$ (\ref{prob-vs-counts-45}). Then, from Eqs. (\ref{cond.prob}) and (\ref{abs-values-45}) one finds all absolute values of parameters $C_i$ and and $C_i(45^\circ)$. And, finally, with known values of $|C_i|$ and $|C_i(45^\circ)|$, one solves numerically Eqs. (\ref{eq-for-phases}) and (\ref{Eq-for-phi2}) and finds the phases $\varphi_1$ and $\varphi_3$ with $\varphi_2=0$. This procedure permits to reconstruct completely all qutrit parameters and to find its degree of entanglement (\ref{conc-squared-simpler}) from experimental coincidence measurements. This procedure might be referred to as an alternative protocol of quantum state tomography of biphoton-based qutrits rather than those one described in \cite{Bogd-04,Bogd-03}.

Note, that if it's known in advance that all parameters $C_i$ are real, the procedure of finding the degree of entanglement from experimental data significantly simplifies. In this case it's more convenient to begin with Eq. (\ref{K-qutrit})  for the inverse Schmidt entanglement parameter, which is easily expressed in terms of single-particle probabilities (\ref{sing-viqa Ci.prob}). For the upper line on the right-hand side of Eq. (\ref{K-qutrit}) we get
\begin{gather}
 \nonumber
 \left(|C_1|^2+\frac{|C_2|^2}{2}\right)^2+\left(|C_3|^2+\frac{|C_2|^2}{2}\right)^2\\
 \label{part1-of-K}
 =w_H^{(s)\,2}+w_V^{(s)\,2}\equiv\frac{1}{2}\left\{1+\left[w_H^{(s)}-w_V^{(s)}\right]^2\right\}.
\end{gather}
On the other hand, with the help of Eqs. (\ref{C123-45})and (\ref{sing-45}), the lower line on the right-hand side of Eq. (\ref{K-qutrit}) can be easily expressed in terms of qutrit parameters and single-particle probabilities in the basis turned for $45^\circ$:
\begin{gather}
 \nonumber
 |C_2^*C_1+C_2C_3^*|^2=C_2^2(C_1+C_3)^2\\
 \label{K-second-part}
  =\frac{\left\{\left[C_1(45^\circ)\right]^2-\left[C_3(45^\circ)\right]^2\right\}^2}{2}
  =\frac{\left(w_{H^{45}}^{(s)}-w_{V^{45}}^{(s)}\right)^2}{2}.
\end{gather}
Altogether this gives
\begin{gather}
 \label{Schmidt-mesurement}
 K^{-1}=
 \frac{1}{2}\left\{1+\left(w_H^{(s)}-w_V^{(s)}\right)^2+\left(w_{H^{(45)}}^{(s)}-w_{V^{(45)}}^{(s)}\right)^2\right\}
\end{gather}
and
\begin{equation}
 \label{C-via-w}
 C=\sqrt{1-\left(w_H^{(s)}-w_V^{(s)}\right)^2-\left(w_{H^{(45)}}^{(s)}-w_{V^{(45)}}^{(s)}\right)^2}.
\end{equation}

\section{Ququarts}

\subsection{Definitions and wave functions}

As mentioned in the Introduction, in the generally accepted treatment, ququarts are considered as two-qubit states. In an abstract form, with unspecified pairs of distinguishable particles, the state vectors of ququarts are taken in the form \cite{Wang}
\begin{equation}
 \label{abcd}
 |\Psi\rangle_{2\,qb}=C_1|00\rangle+C_2|01\rangle
 +C_3|10\rangle+C_4|11\rangle
\end{equation}
with one-qubit single-particle states $|0\rangle$ and $|1\rangle$. Written in the form of two-row columns, the wave function of the state (\ref{abcd}) takes the form
\begin{gather}
 \nonumber
 \Psi_{2\,qb}=C_1\left({1\atop 0}\right)_1\otimes\left({1\atop 0}\right)_2+C_2\left({1\atop 0}\right)_1\otimes\left({0\atop 1}\right)_2\\
 \label{abcd-columns}
 +C_3\left({0\atop 1}\right)_1\otimes\left({1\atop 0}\right)_2+C_4\left({0\atop 1}\right)_1\otimes\left({0\atop 1}\right)_2.
\end{gather}
With this wave function one can easily obtain the reduced density matrix of Ref. \cite{Bogd-06}
\begin{equation}
  \rho_{r\,(2\,qb)}=
  \label{rho-r-Bogd}
  \left(
  {|C_1|^2+|C_2|^2 \quad  C_1C_3^*+C_2C_4^*}\atop
  {C_1^*C_3+C_2^*C_4 \quad |C_3|^2+|C_4|^2}
  \right),
\end{equation}
Schmidt entanglement parameter
\begin{equation}
 K_{2\,qb}=\frac{1}{Tr\big(\rho_{r\,2\,qb}^2\big)}=\frac{1}{1-2|C_1C_4 -C_2C_3|^2},
  \label{Scmidt-2 qb}
\end{equation}
and concurrence of Ref. \cite{Wang}
\begin{equation}
 \label{Conc-Wang}
 C_{2\,qb}=2|C_1C_4-C_2C_3|.
\end{equation}
Though this approach and Eqs. (\ref{rho-r-Bogd})-(\ref{Conc-Wang}) are rather widely used and accepted, we claim that they are inapplicable for biphoton ququarts formed by non-degenerate SPDC pairs. The first objection is the symmetry. The wave function (\ref{abcd-columns}) is asymmetric with respect to the permutation of particle variables $1\rightleftharpoons 2$, and this is strictly forbidden for any biphoton wave functions. A simple symmetrization of the wave function (\ref{abcd-columns}) is insufficient and, actually, it does not help much, because it reduces  the wave function (\ref{abcd-columns}) to the qutrit's wave function (\ref{qutrit-general}), with $C_2+C_3$ playing the role of $C_2$ in Eq.(\ref{qutrit-general}) and  with all peculiarities of ququarts completely lost.

The second objection concerns photon degrees of freedom and dimensionality of the Hilbert space. In contrast to the traditional treatment, biphoton ququarts formed by non-degenerate SPDC pairs of photons have a higher dimensionality than qutrits. Single-photon states from which the biphoton ququarts are constructed are not qubits and they form the Hilbert space of a dimensionality $d=4$, rather than $d=2$ occurring in the case of qutrits. Indeed, if for example, SPDC photons in pairs have two different frequencies $\omega_h$ (high) and $\omega_l$ (low), $\omega_h>\omega_l$, one cannot say for sure, which of two photons gets a higher or lower frequency. Hence, the photon frequency $\omega$ becomes a second photon variable, additional to polarization and taking two values,  $\omega_h$ (high) and $\omega_l$. In other words, each photon has now two degrees of freedom, polarization and frequency \cite{Bu-Ryt}. Together, they make four states which can be occupied by each  photon, or four combined polarization-frequency modes, $Hh,\,Hl,\,Vh,\,{\rm and}\, Vl$ (instead of two polarization modes $H\,{\rm and}\,V$ in the case of degenerate photons). Four modes correspond to the dimensionality of the single-photon Hilbert space $d=4$, in contrast to $d=2$ and two modes in the cases of qutrits and of the traditional two-qubit model of ququarts. The four single-photon polarization-frequency state vectors are now given by
\begin{equation}
 \label{qqrt-single-st-vec}
 a_{Hh}^\dag|0\rangle,\;a_{Hl}^\dag|0\rangle,\;a_{Vh}^\dag|0\rangle,\;a_{Vl}^\dag|0\rangle.
\end{equation}
The corresponding single-photon wave functions, describing states with two degrees of freedom, have two factors, depending on polarization ($\sigma$) and frequency ($\omega$) variables
\begin{gather}
 \nonumber
\delta_{\sigma,H}\delta_{\omega, \omega_h}\equiv\left(1\atop 0\right)^{pol}\otimes\left(1\atop 0\right)^\omega=\left({1\atop 0}\atop{0\atop 0}\right),\quad\quad\\
 \nonumber
 \delta_{\sigma,H}\delta_{\omega, \omega_l}\equiv\left(1\atop 0\right)^{pol}\otimes\left(0\atop 1\right)^\omega\equiv\left({0\atop 1}\atop{0\atop 0}\right), \quad\quad\\
 \nonumber
 \delta_{\sigma,V}\delta_{\omega, \omega_h}\equiv\left(0\atop 1\right)^{pol}\otimes\left(1\atop 0\right)^\omega=\left({0\atop 0}\atop{1\atop 0}\right),\quad\quad\\
 \label{columns-qqrt}
 \delta_{\sigma,V}\delta_{\omega, \omega_l}\equiv\left(0\atop 1\right)^{pol}\otimes\left(0\atop 1\right)^\omega=\left({0\atop 0}\atop{0\atop 1}\right),
\end{gather}
The superscripts $^{pol}$ and $^\omega$ are used here for differentiating the polarization- and frequency- parts of the two-variable wave functions written in the form of direct products of two-row columns.

As dimensionality of the single-photon Hilbert space is $d=4$, these states are $d$=4-qudits, rather than  qubits. Then, as well as qutrits are two-qubit states, the biphoton ququarts are two-qudit states.  Their basis state vectors are given by all possible products of two different creation operators times the vacuum state. But, if we assume that two photons in each non-degenerate SPDC pair definitely have different frequencies, the products of two creation operators corresponding to coinciding frequencies have to be excluded to give finally only four two-photon basis state vectors
\begin{align}
 \nonumber
 |\Psi_{HH}^{(qqrt)}\rangle=\,a_{Hh}^\dag a_{Hl}^\dag|0\rangle,\quad |\Psi_{HV}^{(qqrt)}\rangle=\,a_{Hh}^\dag a_{Vl}^\dag|0\rangle,\\
 \label{qqrt-two-phot-st-vec}
 |\Psi_{VH}^{(qqrt)}\rangle=\,a_{Vh}^\dag a_{Hl}^\dag|0\rangle,\quad |\Psi_{VV}^{(qqrt)}\rangle=\,a_{Vh}^\dag a_{Vl}^\dag|0\rangle,
\end{align}
where, of course, all creation operators commute with each other. The basis wave functions corresponding to the state vectors (\ref{qqrt-two-phot-st-vec}) are obtained with the help of general rules of quantum electrodynamics (\ref{Schweber}):
\begin{gather}
 \nonumber
 \Psi^{(qqrt)}_{HH}=\left({1\atop 0}\right)_1^{pol}\otimes\left({1\atop 0}\right)_2^{pol}\\
 \nonumber
 \otimes\frac{1}{\sqrt{2}}\left[\left({1\atop 0}\right)_1^\omega\otimes\left({0\atop 1}\right)_2^\omega
 +\left({0\atop 1}\right)_1^\omega\otimes\left({1\atop 0}\right)_2^\omega\right]\\
 \label{HH-4}
 \equiv\frac{1}{\sqrt{2}}\left\{
 \left({1\atop 0}\atop{0\atop 0}\right)_1\otimes\left({0\atop 1}\atop{0\atop 0}\right)_2+\left({0\atop 1}\atop{0\atop 0}\right)_1\otimes\left({1\atop 0}\atop{0\atop 0}\right)_2\right\},
\end{gather}

\begin{gather}
 \nonumber
 \Psi^{(qqrt)}_{HV}=\frac{1}{\sqrt{2}}\left[\left({1\atop 0}\right)_1^{pol}\otimes\left({0\atop 1}\right)_2^{pol}\otimes\left({1\atop 0}\right)_1^\omega\otimes\left({0\atop 1}\right)_2^\omega\right.\\
 \nonumber
 +\left.\left({0\atop 1}\right)_1^{pol}\otimes\left({1\atop 0}\right)_2^{pol}\otimes
 \left({0\atop 1}\right)_1^\omega\otimes\left({1\atop 0}\right)_2^\omega\right]\\
 \label{HV-4}
 \equiv\frac{1}{\sqrt{2}}\left\{
 \left({1\atop 0}\atop{0\atop 0}\right)_1\otimes\left({0\atop 0}\atop{0\atop 1}\right)_2+\left({0\atop 0}\atop{0\atop 1}\right)_1\otimes\left({1\atop 0}\atop{0\atop 0}\right)_2
 \right\},
\end{gather}

\begin{gather}
 \nonumber
 \Psi^{(qqrt)}_{VH}=\frac{1}{\sqrt{2}}\left[\left({0\atop 1}\right)_1^{pol}\otimes\left({1\atop 0}\right)_2^{pol}\otimes\left({1\atop 0}\right)_1^\omega\otimes\left({0\atop 1}\right)_2^\omega\right.\\
 \nonumber
 +\left.\left({1\atop 0}\right)_1^{pol}\otimes\left({0\atop 1}\right)_2^{pol}
 \left({0\atop 1}\right)_1^\omega\otimes\left({1\atop 0}\right)_2^\omega\right]\\
 \label{VH-4}
 \equiv\frac{1}{\sqrt{2}}\left\{
 \left({0\atop 0}\atop{1\atop 0}\right)_1\otimes\left({0\atop 1}\atop{0\atop 0}\right)_2+\left({0\atop 1}\atop{0\atop 0}\right)_1\otimes\left({0\atop 0}\atop{1\atop 0}\right)_2
 \right\},
\end{gather}

\begin{gather}
 \nonumber
 \Psi^{(qqrt)}_{VV}=\left({0\atop 1}\right)_1^{pol}\otimes\left({0\atop 1}\right)_2^{pol}\\
 \nonumber
 \otimes\frac{1}{\sqrt{2}}\left[\left({1\atop 0}\right)_1^\omega\otimes\left({0\atop 1}\right)_2^\omega
 +\left({0\atop 1}\right)_1^\omega\otimes\left({1\atop 0}\right)_2^\omega\right]\\
 \label{VV-4}
 \equiv\frac{1}{\sqrt{2}}\left\{
 \left({0\atop 0}\atop{1\atop 0}\right)_1\otimes\left({0\atop 0}\atop{0\atop 1}\right)_2+\left({0\atop 0}\atop{0\atop 1}\right)_1\otimes\left({0\atop 0}\atop{1\atop 0}\right)_2
 \right\}.
\end{gather}
In a general form, the state vector and wave function of the biphoton ququart are given by superpositions of four basis state vectors (\ref{qqrt-two-phot-st-vec}) and four basis wave functions (\ref{HH-4})-(\ref{VV-4})
 \begin{gather}
 \nonumber
 |\Psi\rangle^{(qqrt)}=C_1|\Psi_{HH}^{(qqrt)}\rangle+C_2|\Psi_{HV}^{(qqrt)}\rangle\\
 \label{state-vector-ququart}
 +C_3|\Psi_{VH}^{(qqrt)}\rangle+C_4|\Psi_{VV}^{(qqrt)}\rangle
\end{gather}
and
\begin{equation}
 \label{ququart}
 \Psi^{(qqrt)}=C_1\Psi^{(qqrt)}_{HH}+C_2\Psi^{(qqrt)}_{HV}+C_3\Psi^{(qqrt)}_{VH}+C_4\Psi^{(qqrt)}_{VV}.
\end{equation}
As it should be, both all basis wave functions (\ref{HH-4})-(\ref{VV-4}) and the general ququart wave function (\ref{ququart}) are symmetric with respect to the particle permutations $1\rightleftharpoons2$. Also these expressions take into account properly a higher dimensionality of the biphoton ququarts compared to qutrits and compared to the traditional simplified vision of biphoton ququarts based on Eqs. (\ref{abcd})-(\ref{abcd-columns}).

Once again, dimensionality of the one-photon Hilbert for non-degenerate photons is $d=4$. The dimensionality of the biphoton Hilbert space is $D=d^2=16$. The question is why do ququarts are characterized by only 4 rather than 16 basis wave functions? A general answer is because of some restrictions, owing to which 12 basis wave functions are excluded. One of these restrictions was mentioned above: we assume that frequencies of photons are always different. This excludes states $|h, \sigma;h, \sigma^\prime\rangle$ and $|l, \sigma;l, \sigma^\prime\rangle$, which give rise to 6 symmetric wave functions. Their exclusion does not have fundamental reasons and is related rather with the most often met experimental conditions. The second, deeply fundamental restriction is symmetry. Six remaining and excluded wave functions are antisymmetric, and they cannot be realized with photons at all. Thus, in ququarts all 12 excluded basis wave functions either are missing in all possible superpositions or can be made present symbolically but with obligatory zero coefficients, like the only antisymmetric wave function in the qutrit wave function (\ref{Bell-exp-prime}). As well as in the case of qutrits, the existence of ququarts with only 4 basis wave functions is crucially related to the requirement of symmetry of all biphoton wave functions.

\subsection{Degree of entanglement}

As known \cite{Rungta}, in the case of pure two-particle states the dimensionality of the single-particle Hilbert space $d$ determines directly the maximal achievable value of the Schmidt entanglement parameter, $K_{\max}=d$. Consequently, a two times higher dimensionality of the double-qubit Hilbert space (compared to qubits) doubles the maximal achievable degree of entanglement of ququarts if it's evaluated by the Schmidt entanglement parameter, $K_{\max}=4$ (to be compared with $K_{\max}=2$ in the case of qutrits). To find the entanglement quantifiers of ququarts in a general case, following to the standard procedure, we have to start with finding the density matrix $\rho^{(qqrt)}$, corresponding to the wave function $\Psi^{(qqrt)}$ (\ref{ququart}). Dimensionality of $\rho^{(qqrt)}$ is $16\times 16$ which is too large to be shown explicitly. But the reduced density matrix $\rho_r^{(qqrt)}$ is much more compact, its dimensionality is $4\times 4$ and, explicitly, it is given by
\begin{gather}
 \nonumber
 \rho_r^{(qqrt)}=\\
 \label{reduced-quqv}
 \textstyle\frac{1}{2}
 {\tiny
 \left(
 \begin{matrix}
  |C_1|^2+|C_2|^2 & 0 & C_1C_3^*+C_2C_4^* & 0\\
 0 & |C_1|^2+|C_3|^2 & 0 & C_1C_2^*+C_3C_4^*\\
 C_1^*C_3+C_2^*C_4 & 0 & |C_3|^2+|C_4|^2 & 0\\
 0 & C_1^*C_2+C_3^*C_4 & 0 & |C_2|^2+|C_4|^2
 \end{matrix}\right)
 }
\end{gather}
The derived expression for the $4\times4$ reduced density matrix (\ref{reduced-quqv}) differs significantly from the $2\times 2$ density matrix of Ref. \cite{Bogd-06} where ququarts were considered as two-qubit states. As explained above, the key reasons of this and other differences between this paper and \cite{Bogd-06} and others are in the symmetry of wave functions and in a higher dimensionality of ququarts compared to qutrits properly taken into account in Eqs. (\ref{HH-4})-(\ref{VV-4}) and (\ref{ququart}).

By calculating the squared reduced density matrix $\big(\rho_r^{(qqrt)}\big)^2$ and its trace, we find that the Schmidt entanglement parameter of ququarts is simply twice larger than that obtained in the two-qubit model (\ref{Scmidt-2 qb})\begin{equation}
 K=\frac{2}{1-2|C_1C_4 -C_2C_3|^2}=2K_{2\,qb}.
  \label{Scmidt-general-ququart}
\end{equation}
As for the concurrence, rigorously, its original Wootters' definition
\cite{Wootters} is invalid for two-qudit states with $d>2$. But instead,
one can use the so called {``\it I}-concurrence$"$ \cite{Rungta}, by
definition, determined via the Schmidt entanglement parameter $K$ as
$C_I=\sqrt{2(1-K^{-1})}$. Defined in this way, the  {\it
I}-concurrence does not provide any new information about
entanglement of ququarts compared to that provided by the Schmidt entanglement
parameter $K$. But the  {\it I}-concurrence can be useful for comparison with other entanglement quantifiers,
and for comparison with results of the two-qubit model of ququarts . In a general case Eq.
(\ref{Scmidt-general-ququart}) yields
\begin{gather}
 C_I=\sqrt{1+2|C_1C_4 -C_2C_3|^2}.
 \label{Conc-general-ququart}
\end{gather}
As it follows from Eqs. (\ref{Scmidt-general-ququart}) and (\ref{Conc-general-ququart}), the maximal achievable values of $K$ and $C_I$ for ququarts are $K_{\max}=4$ and $C_{I\,\max}=\sqrt{3/2}$, in agreement with the general expectations for qudits $K_{\max}=d$ and $C_{I\,\max}=\sqrt{2(1-d^{-1})}$ \cite{Rungta}. The minimal values of the Schmidt entanglement parameter and $I$-concurrence are achieved when $C_1C_4=C_2C_3$, and in this case $K=K_{\min}=2$ and  $C_I=C_{I\,\min}=1$. This means that all biphoton ququarts are entangled and nonfactorable (in contrast to this, earlier \cite{Bogd-06}, in a two-qubit model, it was assumed that at $C_1C_4 =C_2C_3$ ququarts are factorable). In particular,  all basis states of ququarts (\ref{HH-4})-(\ref{VV-4}) are minimally entangled. It may be interesting to note that the nature of entanglement of different basis states is different. The wave functions $\Psi_{HH}^{(qqrt)}$ and $\Psi_{VV}^{(qqrt)}$ are seen to be factorized for polarization and frequency parts, and only their frequency parts are entangled, i.e., the states $\Psi_{HH}^{(qqrt)}$ and $\Psi_{VV}^{(qqrt)}$ have only a purely frequency entanglement. On the other hand, the wave functions $\Psi_{HV}^{(qqrt)}$ and $\Psi_{VH}^{(qqrt)}$ are not factorized for polarization and frequency parts, and in these cases we have a nonseparable frequency-polarization entanglement.

Some examples of maximally entangled ququarts are
\begin{gather}
 \nonumber
 \textstyle\frac{1}{\sqrt{2}}\left(\Psi_{HH}^{(qqrt)}\pm\Psi_{VV}^{(qqrt)}\right)\, \left(C_1=\pm C_4=\frac{1}{\sqrt{2}},\,C_3=C_2=0\right),\\
 \nonumber
 \textstyle\frac{1}{\sqrt{2}}\left(\Psi_{HV}^{(qqrt)}\pm\Psi_{VH}^{(qqrt)}\right)\, \left(C_1=C_4=0,\,C_2=\pm C_3=\frac{1}{\sqrt{2}}\right),\\
 \label{max-ent-4}
 \textstyle\frac{1}{2}\left(\Psi_{HH}^{(qqrt)}+\Psi_{HV}^{(qqrt)}+\Psi_{VH}^{(qqrt)}-\Psi_{VV}^{(qqrt)}\right),
\end{gather}
and all functions similar to the last one but with the only sign ``minus$"$ located in front of other terms, $\Psi_{HH}^{(qqrt)}$, $\Psi_{HV}^{(qqrt)}$ or $\Psi_{VH}^{(qqrt)}$.

\begin{figure}[h]
\centering\includegraphics[width=7cm]{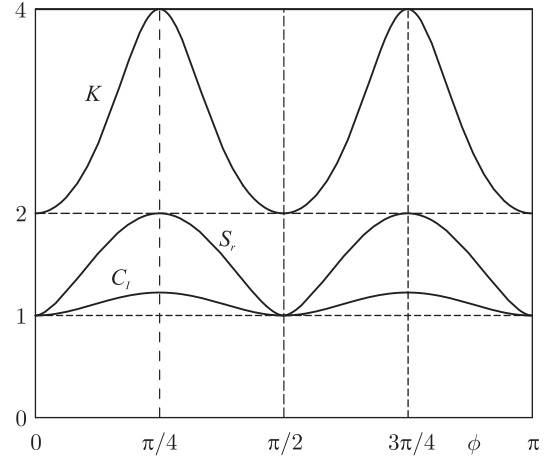}
\caption{{\protect\footnotesize {The Schmidt entanglement parameter $K(\phi)$,
{\it I}-concurrence $C_I(\phi)$, and subsystem
entropy  $S_{r}(\phi)$ for the ququart characterized by the wave
function $\Psi_\phi$ (\ref{example}).}}}\label{Fig4}
\end{figure}

To see explicitly how the degree of entanglement of ququarts changes from $K_{\min}=2$ to $K_{\max}=4$, let us consider, for example, the case $C_1=\cos\phi,\,C_4=\sin\phi,\,C_3=C_2=0$, i.e., the wave function
\begin{equation}
 \label{example}
 \Psi_\phi=\cos\phi\Psi_{HH}^{(qqrt)}+\sin\phi\Psi_{VV}^{(qqrt)}.
\end{equation}
For this state Eqs. (\ref{Scmidt-general-ququart}) and (\ref{Conc-general-ququart}) give $K(\phi)=4/(1+\cos^22\phi)$ and $C_I(\phi)=\sqrt{1+\frac{1}{2}\sin^22\phi}$. These functions are plotted in Fig. \ref{Fig4} together with the subsystem entropy $S_r(\phi)=-\sum_{i=1}^4
\lambda_i\log_2\lambda_i$, where $\lambda_i$ are eigenvalues of
the reduced density matrix $\rho_r^{(qqrt)}$ (\ref{reduced-quqv}). In
the case $\Psi^{(qqrt)}=\Psi_\phi$ (\ref{example}) the matrix
$\rho_{r\,\phi}^{(qqrt)}$ is very simple
\begin{equation}
 \label{very simple}
\rho_{r\,\phi}^{(qqrt)}=\left(\begin{matrix}\frac{1}{2}\cos^2\phi &
0&0&0\\0&\frac{1}{2}\cos^2\phi
&0&0\\0&0&\frac{1}{2}\sin^2\phi&0\\0&0&0&\frac{1}{2}\sin^2\phi\end{matrix}\right).
\end{equation}
Evidently, its eigenvalues are
$\lambda_1=\lambda_2=\frac{1}{2}\cos^2\phi$ and
$\lambda_3=\lambda_4=\frac{1}{2}\sin^2\phi$, which yields
\begin{equation}
 \label{entr-ququart}
 S_r(\phi)= 1-2\left(\cos^2\phi\log_2|\cos\phi|+\sin^2\phi\log_2|\sin\phi|\right).
\end{equation}

Eigenvalues of the reduced density matrix $\rho_{r\,\phi}^{(qqrt)}$ are twice
degenerate $\lambda_1=\lambda_2$ and $\lambda_3=\lambda_4$. If
none of them equals zero, the Schmidt decomposition of the wave function $\Psi_\phi$
contains four terms of Schmidt-mode products. However, at $\phi=0$, when
the entropy $S_r(\phi)$ is minimal, two of these eigenvalues turn
zero, $\lambda_3=\lambda_4=0$, and two Schmidt-mode products in the Schmidt
decomposition disappear. It is the reason why in this case the degree of entanglement
is minimal. However, the remaining two terms in the Schmidt
decomposition indicate that in the case $\phi=0$ the state $\Psi_\phi$ is entangled and non-factorable.
The degree of entanglement of the state $\Psi_\phi$ is maximal when $\lambda_1=\lambda_2=\lambda_3=\lambda_4=1/4$, i.e., when all four Schmidt-mode products enter the Schmidt decomposition with equal weights, and this occurs at $\phi=45^{\circ}$ and $\phi=135^{\circ}$.

\begin{figure}[h]
\centering\includegraphics[width=7cm]{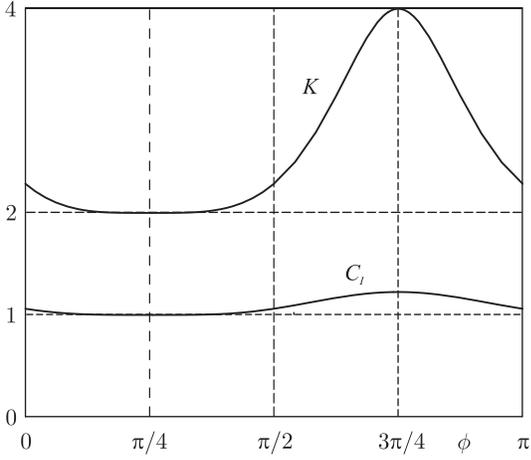}
\caption{{\protect\footnotesize {The Schmidt entanglement parameter $K(\phi)$,
and {\it I}-concurrence $C_I(\phi)$ for the ququart characterized by the wave
function $\Psi_\phi$ (\ref{example-2}).}}}\label{Fig5}
\end{figure}

In a general case, ququarts are multi-parametric objects, and it's rather difficult to show all their features in a limited amount of pictures. However, it may be interesting to show one example more, when the ququart wave function has the form
\begin{equation}
 \label{example-2}
 \Psi_\phi^\prime=\frac{\cos\phi\Psi_{HH}^{(qqrt)}+\sin\phi\Psi_{VV}^{(qqrt)}}{\sqrt{2}}
 +\frac{\Psi_{HV}^{(qqrt)}+\Psi_{VH}^{(qqrt)}}{2},
\end{equation}
for which $C_1=\cos\phi/\sqrt{2}$, $C_4=\sin\phi/\sqrt{2}$, and $C_2=C_3=1/2$.
The Schmidt entanglement parameter $K(\phi)$ and $I$-concurrence $C(\phi)$ of the state (\ref{example-2}) are shown in Fig. \ref{Fig5}.
Asymmetry of these curves with respect to the substitution $\phi\rightarrow\pi-\phi$ indicates sensitivity of the degree of entanglement with respect to the sign of the product $C_1C_3$, i.e., to phases of these real constants.

To conclude this subsection, let us describe here one example more of a rather peculiar polarization-angular ququart obtainable from qutrits with the help of a beam splitter and post-selection. As mentioned earlier, polarization-frequency and polarization-angular ququarts are equivalent. Nondegeneracy of photons can be provided by different direction of photons with equal frequencies in the noncollinear SPDC process. Another way of providing noncollinear propagation of originally collinear and frequency-degenerate photons is in using a nonselective beam splitter. As discussed above in Section {\bf 6A}, after BS photons acquire new variables, $\theta_1$ and $\theta_2$, which can take one of two values each, $0^\circ$ and $90^\circ$. The combined polarization-angular wave function after BS is given by Eq. (\ref{bs}). This wave function is not yet a complete analog of ququarts considered above because of unsplit pairs in Fig. \ref{Fig3}. Mathematically they correspond to terms in Eq. (\ref{bs}) proportional to $\delta_{\theta_1,0}\delta_{\theta_2,0}$ and $\delta_{\theta_1,90^\circ}\delta_{\theta_2,90^\circ}$. In the case of frequency nondegenerate biphoton states similar terms would describe both photons with high or both with low frequencies, which we have excluded (see the discussion in the previous Subsection). Owing to these terms with coinciding angles $\theta_1$ and $\theta_2$, as mentioned above, the angular part of the combined wave function in Eq. (\ref{bs}) does not provide any additional entanglement to the polarization part $\Psi$ (qutrit). However, if we make a post-selection by killing (eliminating) in some way unsplit pairs in both channels I and II of Fig. \ref {Fig3}, we get a true polarization-angular ququart with the wave function of the form
\begin{equation}
 \label{qtr-to-qqrt}
 \Psi^{qqrt}=\Psi^{qtr}(\sigma,\sigma^\prime)\frac{1}{\sqrt{2}} [\delta_{\theta_1,0}\delta_{\theta_2,90^\circ}+\delta_{\theta_1,90^\circ}\delta_{\theta_2,0}],
\end{equation}
where $\Psi^{qtr}(\sigma,\sigma^\prime)$ is an arbitrary qutrit wave function and, as usual, $\sigma$ and $\sigma^\prime$ are polarization variables of photons. Peculiarity of the ququart, characterized by the wave function (\ref{qtr-to-qqrt}) consists in factorization of parts depending on polarization and angular variables. For this reason the total Schmidt entanglement parameter of the state (\ref{qtr-to-qqrt}) appears to be factorable too:
\begin{equation}
 \label{K-factored}
 K^{qqrt}=K^{qtr}\times K^{angle},
\end{equation}
where $K^{qtr}$ and $K^{angle}$ are the Schmidt entanglement parameters of the qutrit $\Psi^{qtr}$ and of the angular part of the wave function $\Psi^{qqrt}$, $\frac{1}{\sqrt{2}} [\delta_{\theta_1,0}\delta_{\theta_2,90^\circ}+\delta_{\theta_1,90^\circ}\delta_{\theta_2,0}]$. Actually, evidently enough, the Schmidt entanglement parameter of the state, characterized by the last wave function equals two, $K^{angle}=2$. For this reason Eq. (\ref{K-factored}) takes the simplest form:  $K^{qqrt}=2 K^{qtr}$, i.e., in terms of the Schmidt entanglement parameter, the post-selection after BS doubles the degree of entanglement of the original qutrit. Though interesting enough, this last result and factorization of the Schmidt entanglement parameter described by Eq. (\ref{K-factored}) are specific features of the ququart of the form (\ref{qtr-to-qqrt}). In a general case the degree of entanglement of ququarts, characterized by any entanglement quantifiers, is unseparable for angular/frequency and polarization parts.

\subsection{Measurements}

A possible way for measuring directly parameters of ququarts and, in particular, their degree of entanglement is similar to that described above for qutrits and illustrated by Fig. \ref{Fig3}: the original biphoton beam has to be split for two channels by a nonselective beam splitter and a full set of coincidence measurements has to be done in the $xy$ and $x^\prime y^\prime$ bases of Fig. \ref{Fig3}$b$. In difference with qutrits, in addition to polarizers one has to install frequency filters in front of detectors to count amounts of photons in given frequency-polarization  $Hh$, $Hl$, $Vh$, and $Vl$. As in ququarts all basis biphoton states are of the type $a_i^\dag a_j^\dag |0\rangle$ with $i\neq j$ (where $i$ and $j$ numerate above indicated modes), distribution of photons of split pairs in channels I and II corresponds to the type, shown in Fig. \ref{Fig3}$b$. For this reason, if  $N^{tot}_{i,j}$ are total amounts of generated pairs with photons in modes $i$ and $j$, the amounts of the corresponding coincidence detector counts are determined by equations similar to that of the last formula in Eq. (\ref{counts-pairs}):
\begin{equation}
 \label{counts-pairs-qqrt}
 \left.N^{d\,(c)}_j\right|_j=\left.N^{d\,(c)}_j\right|_i=\frac{\eta}{4}N^{tot}_{i,j},
\end{equation}
with $\eta$ denoting the efficiency of detectors. The full set of data on coincidence amounts of counts can be used further to find conditional probabilities for a photon ``1$"$ to be in some mode $i$ under the condition that the second photon of the same pair is in some other mode $j$ (similarly to Eq. (\ref{prob-vs-counts}))
\begin{equation}
 \left.w_i^{(c)}\right|_{j}=\frac{\left.N^{d\,(c)}_i\right|_{j}}
 {\sum_{i,j}\left.N_i^{d\,(c)}\right|_j}
  \label{prob-vs-counts-qqrt}
\end{equation}
with the normalization condition $\sum_{i,j}\left.w_i^{(c)}\right|_{j}=1$.

On the other hand, the conditional probabilities are determined by diagonal elements of the full density matrix with respect to both polarization and frequency variables and of both photons. Directly from expressions (\ref{HH-4})-(\ref{VV-4}) for the basis wave functions of ququarts, from the general expression (\ref{ququart}) for the ququart wave function, and from the definition of the density matrix $\rho=\Psi\Psi^\dag$  we find the following series of relations between the conditional probabilities and the parameters $C_i$ of ququarts
\begin{equation}
 \label{c-vs-w-qqrt}
 \setlength{\extrarowheight}{0.2cm}
 \begin{matrix}
 |C_1|^2=2w_{Hh}^{(c)}\Big|_{Hl}=2w_{Hl}^{(c)}\Big|_{Hh},\\
 |C_2|^2=2w_{Hh}^{(c)}\Big|_{Vl}=2w_{Vl}^{(c)}\Big|_{Hh},\\
 |C_3|^2=2w_{Hl}^{(c)}\Big|_{Vh}=2w_{Vh}^{(c)}\Big|_{Hl},\\
 |C_4|^2=2w_{Vh}^{(c)}\Big|_{Vl}=2w_{Vl}^{(c)}\Big|_{Vh}.
 \end{matrix}
\end{equation}
The same relations (\ref{counts-pairs-qqrt})-(\ref{c-vs-w-qqrt}) between amounts of detector counts, conditional probabilities and expansion coefficients of the ququart wave function can be written also in the $x^\prime y^\prime$ basis turned for $45^\circ$ with respect to the horizontal-vertical basis. Thus, all set of coincidence measurements in two bases can be used to determine all absolute values of ququarts parameters in both bases, $|C_i|^2$ and $|C_i(45^\circ)|^2$. Relations between $C_i(45^\circ)$ and $C_i$ are easily found in the same way as in the case of qutrits, and in the case of ququarts they are given by
\begin{align}
 \label{C-0-45-qqrt}
 \setlength{\extrarowheight}{0.2cm}
 \begin{matrix}
 C_1(45^\circ)=\textstyle\frac{1}{2}\,\left(C_1+C_2+C_3+C_4\right),\\
 C_2(45^\circ)=\textstyle\frac{1}{2}\,\left(-C_1+C_2-C_3+C_4\right),\\
 C_3(45^\circ)=\textstyle\frac{1}{2}\,\left(-C_1-C_2+C_3+C_4\right),\\
 C_4(45^\circ)=\textstyle\frac{1}{2}\,\left(C_1-C_2-C_3+C_4\right).
 \end{matrix}
\end{align}

The squared absolute values of expressions on the left- and right-hand sides of Eqs. (\ref{C-0-45-qqrt}) take the form of equations for the phases $\varphi_{1,2,3,4}$ of the constants $C_i$. For shortening formulas these equations can be written grouped in pairs to give
\begin{gather}
 \nonumber
 |C_1(45^\circ)|^2+|C_2(45^\circ)|^2-\textstyle\frac{1}{2}=\\\
 \label{1+2}
 |C_1||C_3|\cos(\varphi_1-\varphi_3)+|C_2||C_4|\cos(\varphi_2-\varphi_4),
\end{gather}
\begin{gather}
 \nonumber
 |C_1(45^\circ)|^2+|C_3(45^\circ)|^2-\textstyle\frac{1}{2}=\\
 \label{1+3}
 |C_1||C_2|\cos(\varphi_1-\varphi_2)+|C_3||C_4|\cos(\varphi_2-\varphi_4),
\end{gather}
\begin{gather}
 \nonumber
 |C_1(45^\circ)|^2+|C_4(45^\circ)|^2-\textstyle\frac{1}{2}=\\
 \label{1+4}
 +|C_1||C_4|\cos(\varphi_1-\varphi_4)+|C_2||C_3|\cos(\varphi_2-\varphi_3).
\end{gather}
Because of normalization $\sum_i|C_i(45^\circ)|^2=1$, one can get only three independent equations of the type (\ref{1+2})-(\ref{1+4}) from four Eqs. (\ref{C-0-45-qqrt}). The fourth equation for finding four phases $\varphi_{1,2,3,4}$ follows from the above mentioned fact that the phase of the wave wave function $\Psi^{(qqrt)}$ (\ref{ququart}) as a whole does not affect any observable quantities and can be arbitrarily chosen. For this reason one can put some additional condition restriction for phases, e.g., such as
\begin{equation}
 \label{1+2+3+4}
 \varphi_1+\varphi_2+\varphi_3+\varphi_4=0.
\end{equation}
Four equations (\ref{1+2})-(\ref{1+2+3+4}) are sufficient for finding numerically all phases $\varphi_{1,2,3,4}$ as soon as the squared absolute values $|C_i|^2|$ and $|C_i(45^\circ)|^2$  are found from coincidence measurements. Thus the procedure described above provides a full reconstruction of the ququart wave function $\Psi^{(qqrt)}$ (\ref{ququart}) and, in particular, can be used for determining the degree of entanglement. This procedure serves as an alternative protocol of quantum state tomography of ququarts described in the work \cite{Bogd-06}.

Note that, in analogy with the case of qutrits, determination of entanglement quantifiers of ququarts from experimental data is simplified if the parameters $C_{1,2,3,4}$ are known in advance to be real. In this case the key element of Eqs. (\ref{Scmidt-general-ququart}) and (\ref{Conc-general-ququart}) for the Schmidt entanglement parameter and $I$-concurrence, $|C_1C_4-C_2C_3|^2$ can be written as
\begin{equation}
 \label{14-23}
 |C_1C_4-C_2C_3|^2=C_1^2C_4^2+C_2^2C_3^2-2C_1C_4C_2C_3.
\end{equation}
On the other hand, by summing the squared first and last lines of Eqs. (\ref{C-0-45-qqrt}) we get
\begin{gather}
 \label{14+23}
 C_1C_4+C_2C_3=C_1^2(45^\circ)+C_4^2(45^\circ)^2-\textstyle\frac{1}{2}.
\end{gather}
Now, with a simple algebra, we find from Eqs. (\ref{14-23}) and (\ref{14+23}) the following final expression for $|C_1C_4-C_2C_3|^2$
\begin{gather}
 \nonumber
 |C_1C_4-C_2C_3|^2=2\left(|C_1|^2|C_4|^2+|C_2|^2|C_3|^2\right)\\
 \label{14-23-fianl}
 -\left[|C_1(45^\circ)|^2+|C_4(45^\circ)|^2-\textstyle\frac{1}{2}\right]^2,
\end{gather}
where all terms on the right-hand side are expressible via conditional probabilities either in the original ($xy$) basis, or in the basis turned for $45^\circ$. This solves the problem of determining the entanglement quantifiers $K$ or $C_I$.

\section{Discussion}

One of the key elements of the approach of this paper consists in the analysis how does the obligatory symmetry of biphoton wave functions affect the particle entanglement. The conclusion is that the symmetry restrictions on biphoton states imposed by the Boze statistics are found to be important not only for correct calculations of the entanglement quantifiers but also for existence of biphoton qutrits and ququarts as classes of states with, correspondingly, only three and four basis states. Actually, for  biphoton states with dimensionality $D=4$ (qutrits) and $D=16$ (ququarts) one can expect appearance of, correspondingly, four and sixteen basis wave functions. In both cases such formations would be  more complicated than qutrits and ququarts. To return to true qutrits and ququarts, we have to restrict the amount of basis wave functions by excluding some of them, and the exclusion has to result from some physics rather than being made artificially. The key role in such restriction of the amount of basis wave functions belongs to the symmetry requirements. In the case of qutrits the wave function to be excluded is the antisymmetric Bell state $\Psi^-$ (\ref{Bell-fourth}), which cannot arise in any superpositions of purely polarization biphoton wave functions, because its appearance would contradict to the Boze statistics of photons. Alternatively, this fourth antisymmetric basis wave function can be included in superpositions but with obligatory zero coefficient, as in Eq. (\ref{Bell-exp-prime}). In the case of ququarts the situation is even more dramatic. To get four basis wave functions instead of sixteen we have to exclude twelve wave functions. Six of them are excluded because we consider here only SPDC processes with non-coinciding frequencies of photons, and the remaining six basis wave functions are excluded because of symmetry restrictions as antisymmetric ones and, hence, forbidden by the Boze statistics. Thus, indeed, in the case of biphoton ququarts, symmetry restrictions are crucially important for existence of such class of states.

The question that is often raised concerns relations between the fundamental entanglement arising owing to symmetry of wave functions and ``configurational$"$ entanglement arising owing to varying choice of coefficients $C_i$ in superpositions of basis wave functions (\ref{qutrit-general}) and (\ref{ququart}). As seen from our analysis, the entanglement arising owing to symmetry is the intrinsic feature of all basis wave functions of ququarts (\ref{HH-4})-(\ref{VV-4}) and of the qutrit's basis wave function $\Psi_{HV}^{qtr}$ (\ref{HV}), (\ref{columns-HV}). This entanglement is fixed and cannot be changed by any basis transformations. On the other hand, entanglement arising in superpositions of basis functions depends on the coefficients in these superpositions. In this sense one can speak about configurational entanglement as an ``addition$"$ to entanglement arising owing to symmetry of wave functions. But the word ``addition$"$ should not be understood in this case as summing different types of entanglement. In a general case none of the entanglement quantifiers equals to a sum or a product of the same quantifiers, related to symmetry and to superposition of basis wave functions. I.e., the symmetry and configurational entanglement are unseparable from each other and together they determine the degree of entanglement of ququarts and qutrits. Related to this, it's often assumed that entanglement arising owing to symmetry is resourceless whereas the configurational entanglement is resourceful for applications such as protocols of quantum information and quantum communication like quantum teleportation \cite{teleport}, dense coding \cite{DC}, quantum key distribution \cite{Ekert}, etc. Moreover, sometimes this idea is extrapolated to the assumption that entanglement related to symmetry can be ignored at all or ``forgotten$"$. We think that such formulations are excessive. As said above, symmetry and configurational entanglements are unseparable from each other, and it's impossible to distinguish in a total entanglement the resourceless and resourceful parts. On the other hand, it's a rather interesting and important question what changes  in applications of ququarts analyzed earlier in the frame of the two-qubit model when the latter is substituted by the above described two-qudit picture. We hope to return to this manifold of problems elsewhere.

In connection with different opinions on entanglement related to symmetry of indistinguishable particles, it's worth mentioning the formulation by A. Peres \cite{Peres}: ``We must now convince ourselves that this entanglement is not a matter of concern: No quantum prediction, referring to an atom located in our laboratory, is affected by the mere presence of similar atoms in the remote universe$"$. The second part of this statement is absolutely correct: far objects cannot affect any measurements performed solely with a laboratory object. But as for ``not a matter of concern$"$, it depends on for what and under what conditions. For local laboratory measurements it's not a matter of concern, but measurements of the degree of entanglement assume coincidence measurements carried out in both laboratory and remote systems, and they assume a possibility of communication between systems for selecting only coinciding signals. Only if these possibilities occur, we can formulate a question about the degree of entanglement of the laboratory and remote particles. Then, if particles are indistinguishable and if their common state is pure, the entanglement related to symmetry can be a matter of concern for correct determining the total degree of entanglement of a bipartite system. To specify a little bit further these general speculations, let us consider the same example as in \cite{Peres}: two well separated identical one-electron atoms, ``$a"$ and ``$b"$, both in the ground states. Let the two-electron state be pure and let the spin state of two electrons be symmetric and non-entangled,  $\left|\uparrow\uparrow\right\rangle$. Then the coordinate wave function of two electrons is
\begin{gather}
 \nonumber
 \psi_0(x_1,x_2)=\\
 \label{2-ei-wf-gr}
 \frac{\big[\varphi_0(x_1-a)\varphi_0(x_2-b)-\varphi_0(x_2-a)\varphi_0(x_1-b)\big]}{\sqrt{2}},
\end{gather}
where $x_1$ and $x_2$ are positions of atomic electrons, $a$ and $b$ are given positions of atomic nuclei, both $x_{1,2}$ and $a$, $b$ can be vectors. We can assume that atoms are located far enough for the wave functions $\varphi_0(x-a)$ and $\varphi_0(x-b)$ to not overlap, which means that $\int dx \varphi_0^*(x-a)\varphi_0(x-b)=0$. Eq. (\ref{2-ei-wf-gr}) takes into account indistinguishability of electrons and the necessary Fermi-Dirac symmetry requirements $\psi_0(x_1,x_2)=-\psi_0(x_2,x_1)$. Owing to this, we cannot say, which of electrons, ``1$"$ or  ``2$"$ is located near the atom ``$a"$, and which near the atom ``$b"$. To illustrate that, indeed, local measurements with one of two atoms are not affected by antisymmetrization of the wave function, we can consider, e.g., the process of laser excitation solely of an atom ``$a"$. This situation can occur if the laser field is focused around the point $a$ and is zero around $b$. The excited two-electron state with an electron in the atom $a$ being at an excited level, and in the atom $b$ - in the ground state has the wave function of the form:
\begin{gather}
 \nonumber
 \psi_1(x_1,x_2)=\\
 \label{2-ei-wf-ex}
 \frac{\big[\varphi_1(x_1-a)\varphi_0(x_2-b)-\varphi_1(x_2-a)\varphi_0(x_1-b)\big]}{\sqrt{2}},
\end{gather}
where $\varphi_1(x-a)$ is the electron wave function at the excited level. As we don't know which electron is located near the atom ``$a"$, we have to take the atom-light dipole interaction in the form $V\cos\omega t$ with $V=-e\varepsilon_0 (t)[(x_1-a)+(x_2-a))]$, where $\varepsilon_0$ and $\omega$ are the field-strength amplitude and frequency of the exciting field. The rate of transitions found with the help of the Fermi Golden Rule is given by
\begin{gather}
 \nonumber
 {\dot w}= \frac{2\pi}{\hbar}\left|\left\langle\psi_1\left|\frac{V}{2}\right|\psi_0\right\rangle\right|^2\delta(E_1-E_0-\hbar\omega)\\
 \label{FGR}
 \equiv\frac{2\pi}{\hbar}\left|\left\langle\varphi_1(x)
 \left|\frac{e\varepsilon_0x}{2}\right|\varphi_0(x)\right\rangle\right|^2\delta(E_1-E_0-\hbar\omega),
\end{gather}
where $E_0$ and $E_1$ are energies of the ground and excited atomic states. Identity of expressions in the upper and lower lines of Eq. (\ref{FGR}) confirms that for local excitations only in one of two atoms the second atom does not affect the rate of transitions at all and the latter can be found without anisymmetrization of the two-electron wave function, simply with $\psi_0$ (\ref{2-ei-wf-gr}) and $\psi_1$ (\ref{2-ei-wf-ex}) substituted by
\begin{equation}
 \label{non-symm-wf}
 {\tilde\psi}_0=\varphi_0(x_1-a)\varphi_0(x_2-b),\;{\tilde\psi}_1=\varphi_1(x_1-a)\varphi_0(x_2-b).
\end{equation}

However, if we are interested in whether the state of two one-electron atoms in ground states is entangled or not, we have to find some of its entanglement quantifiers. E.g., the Schmidt entanglement parameter in the case of continuous variables $x_1$  and $x_2$ is known \cite{JPB} to be given by
\begin{gather}
 \nonumber
 K^{-1}=\int dx_1dx_2dx_1^\prime dx_2^\prime \,\psi_0(x_1,x_2)\psi_0^*(x_1^\prime,x_2)\\
 \label{two-at-Schm}
 \times \psi_0^*(x_1,x_2^\prime)\psi_0(x_1^\prime,x_2^\prime).
\end{gather}
 For two remote atoms the integral on the right-hand side equals $\frac{1}{2}$, which gives $K=2$ and which means that the state of two remote atoms characterized by a pure two-electron wave function is always entangled. ``Always$"$ means in this case as long as the state $\psi_0(x_1,x_2)$ remains pure, i.e., as long as its purity is not hurt by external factors. Also this means that two atoms have to arise in a pure two-electron state. An example when it's true is the dissociation of a hydrogen molecule by a Fourier-limited light field for two Hydrogen atoms. An opposite example, when the state of electrons in two remote atoms is hardly pure, is when the atoms are obtained from different sources. In particular, the two-electron state hardly is pure and hardly can be described by a wave function when one of them is produced in the Earth laboratory and another somewhere in a remote Universe. For such atoms the quoted above formulation of Ref. \cite{Peres} is absolutely correct: remote objects arising from different sources are non-entangled, and symmetrization or antisymmetrization of their wave functions has no sense because their common state cannot be characterized by a bipartite wave function. But for remote indistinguishable particles in a pure bipartite state their entanglement arising owing to symmetry can occur, can be important and measurable, and is unseparable from the configurational entanglement related to superposition of basis states.

\section{Conclusions}

To summarize, the consideration given above provided a systematic description of biphoton qutrits and ququarts and such their features as symmetry, dimensionality, and entanglement. As both qutrits and ququarts are two-bozon formations, their wave functions are obliged to be symmetric with respect to permutations of photon variables, and this condition is important for correct evaluation of the degree of entanglement. In our analysis the degree of entanglement is evaluated by such entanglement quantifiers as the Schmidt entanglement parameter, concurrence, and subsystem von Neumann entropy. All of them are good for pure bipartite states, they characterize the degree of entanglement in different metrics but, of course, their predictions must be and they are compatible with each other. The Schmidt-mode analysis is also applied and shown to be very fruitful for finding families of non-entangled, factorable qutrits. For qutrits we found explicitly the most general three-parametric families of non-entangled and maximally entangled states. In particular, we showed  that one of the basis state of qutrits, in which photons have different polarization (horizontal and vertical), belongs to the family of maximally entangled qutrits. Some other interesting features of qutrits are analyzed, such as, e.g., anticorrelation of entanglement and polarization, and so on. In the case of biphoton ququarts, even more interesting, the traditional two-qubit model is shown to be invalid. As ququarts are produced by biphoton beams with nondegenerate photons (either in frequencies or in directions of propagation), they have more degrees of freedom than qutrits, e.g., polarization and frequency. In such case frequencies of photons are variables rather than simply some given numbers, and  frequency entanglement has to be taken into account together with polarization entanglement. Thus, biphoton ququarts are shown to be two-qudit rather than two-qubit states, with the dimensionality of the one-photon Hilbert space $d=4$  and dimensionality of the two-photon Hilbert space $D=d^2=16$ (in contrast with $d=2$ and $D=d^2=4$ in the case of qutrits). This new understanding of the physics of biphoton ququarts gave new formulas for their entanglement quantifiers (\ref{Scmidt-general-ququart}) and (\ref{Conc-general-ququart}). One of qualitative consequences of these results is that all ququarts are entangled and unseparable, in contrast to earlier predictions of the two-qubit model.
For both qutrits and ququarts, schemes of their complete reconstruction from experimental data are suggested. The schemes are based on using a nonselective beam splitter and carrying out full sets of coincidence photon-counting measurements in the usual horizontal-vertical basis and in the basis turned for $45^\circ$.
\begin{acknowledgments}
The work is supported partially by the RFBR grants 08-02-01404 and 10-02-90036.
\end{acknowledgments}

\end{document}